\documentclass[%
 reprint,
superscriptaddress,
 amsmath,amssymb,
 aps,
]{revtex4-2}

\usepackage{graphicx}
\usepackage{dcolumn}
\usepackage{bm}
\usepackage{color}

\newcommand{\ave}[1]{{\langle #1\rangle}}
\newcommand{\ii}{ {\rm i} }

\usepackage{dsfont,bbm,xcolor}
\usepackage{enumitem}
\usepackage{hyperref}
\begin{document}

\preprint{APS/123-QED}

\title{Tunable Non-equilibrium Phase Transitions between Spatial and Temporal Order through Dissipation}

\author{Zhao Zhang}\affiliation{Institute for Quantum Electronics, Eidgenössische Technische Hochschule Zürich, Otto-Stern-Weg 1, CH-8093 Zurich, Switzerland}
\author{Davide Dreon}\affiliation{Institute for Quantum Electronics, Eidgenössische Technische Hochschule Zürich, Otto-Stern-Weg 1, CH-8093 Zurich, Switzerland}
\author{Tilman Esslinger}\affiliation{Institute for Quantum Electronics, Eidgenössische Technische Hochschule Zürich, Otto-Stern-Weg 1, CH-8093 Zurich, Switzerland}
\author{Dieter Jaksch}\affiliation{Institut für Laserphysik, Universität Hamburg, 22761 Hamburg, Germany}\affiliation{Clarendon Laboratory, University of Oxford, Parks Road, Oxford OX1 3PU, United Kingdom}
\author{Berislav Buca} \email{berislav.buca@physics.ox.ac.uk} \affiliation{Clarendon Laboratory, University of Oxford, Parks Road, Oxford OX1 3PU, United Kingdom} 
\author{Tobias Donner}\affiliation{Institute for Quantum Electronics, Eidgenössische Technische Hochschule Zürich, Otto-Stern-Weg 1, CH-8093 Zurich, Switzerland}

\date{\today}
\begin{abstract}
We propose an experiment with a driven quantum gas coupled to a  dissipative optical cavity that realizes a novel kind of far-from-equilibrium phase transition between spatial and temporal order. The control parameter of the transition is the detuning between the drive frequency and the cavity resonance. For negative detunings, the system features a spatially ordered phase, while positive detunings lead to a  phase with both spatial order and persistent oscillations, which we call dissipative spatio-temporal lattice. We give numerical and analytical evidence for this superradiant  phase transition and show that the spatio-temporal lattice originates from cavity dissipation. In both regimes the atoms are subject to an accelerated transport, either via a uniform acceleration or via abrupt transitions to higher momentum states. Our work provides perspectives for temporal phases of matter that are not possible at equilibrium. 
\end{abstract}


\maketitle

\emph{Introduction—} Crystallization is usually known as a phase transition, where a system  of many particles at initially random positions self-arranges to acquire order in space \cite{cardy_1996}. Recently, also the concept of temporal crystallization has been introduced \cite{Wilczek}, where a system of many particles enters a state with repetitive motion, that is, order in time. Since space and time are the two fundamental continuous degrees of freedom required to describe nature and its evolution, spatial and temporal order  are also the two types of fundamentally possible crystals. While both types of crystals and their phase transitions have been investigated, the possibility of a phase transition between these two types of order has so far not been discussed. In this work we present a proposal for a system with a phase diagram featuring such a transition between spatial and temporal order in a dissipative setting. Counter-intuitively, dissipation can be engineered to induce persistent oscillations and temporal crystallization in quantum systems~\cite{Dissipative1,Orazio,Nunnenkamp,Seibold,Seifert,JuanJose2,Fabrizio,Lesanovsky,Jamir3,chen2021collectively,seibold2021quantum,Buca_2019,Booker_2020,Fazio,sarkar2021signatures}. Quantum optical setups provide promising platforms for engineering quantum phases of matter through dissipation~\cite{Cavity2,cavityreview,Cavity4,Piazza1,Hadiseh,Keeling,Keeling2}. Transitions to steady states of self-ordered crystals in space have been broadly explored~\cite{Cavity1,Cavity2,cavityreview}, and in particular both continuous \cite{Buca_2019,PhysRevLett.123.260401,Lledo1} and discrete dissipative time crystals \cite{Jamir1,Jamir2,Chinzei,Cosme1} have been recently realised as non-stationary states in such systems~\cite{Dogra1496,dissipativeTCobs,DissipativeContObs,dreon2021self}. However, so far transitions between spatial and temporal order have not been studied.

\begin{figure}[ht]
\centering
\includegraphics[scale=0.37]{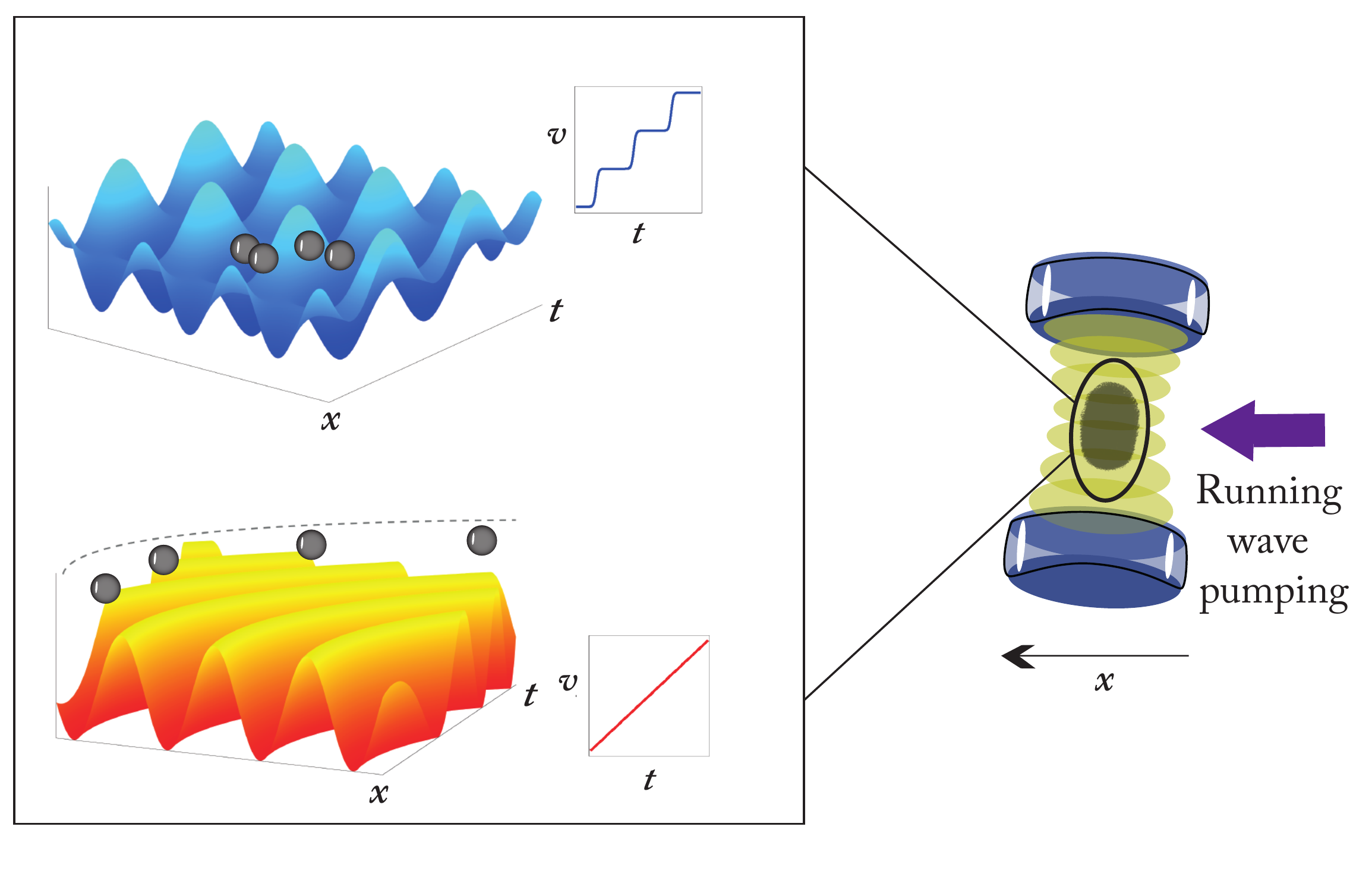}
\caption{A Bose-Einstein condensate confined to one-dimension and trapped inside a lossy optical cavity is illuminated with a running wave laser field (right). Depending on the detuning between cavity and laser frequency, the system enters either a phase where the atoms constantly accelerate by a co-moving optical lattice (red structure), or a phase where they are pumped in a periodic fashion to increasingly higher momentum states, forming a spatio-temporal lattice. A phase transition between these two regimes can be explored by changing the detuning. }
\label{fig:Intro}
\end{figure}

Here we present a proposal for realizing such a novel kind of phase transition between a phase with spatial order alone (spatial lattice) and a phase with simultaneous crystallization in space and time (dissipative spatio-temporal lattice). In our envisioned setup, a bosonic quantum gas is dispersively coupled to a dissipative optical cavity and pumped with a transverse laser field (see Fig.~\ref{fig:Intro}). In contrast with similar setups~\cite{HemmerichPNAS,LevNatComms,baumann2010dicke,ZhangScience,dissipativeTCobs,DissipativeContObs}, we use a running-wave pumping beam instead of a standing-wave beam, which unbalances the forward and backward photon scattering processes. Since the laser frequency is far detuned from the atomic resonance, the atoms act as scatterers at which photons from the pump field can be scattered into the cavity mode and vice versa. We consider a one-dimensional situation along the propagation direction of the pumping field, which can be effectively realized using a $\lambda$-periodic lattice along the cavity axis. 
As the initial state we assume a coherent momentum state of the atoms making the setup far-from-equilibrium. Employing the pumping strength $\eta_p$ and the detuning $\Delta_c=\omega_p - \omega_c$ between the transverse laser frequency $\omega_p$ and the cavity resonance $\omega_c$ as control parameters with the dynamic dispersive shift \cite{mivehvar2021cavity} being absent in one-dimension, we identify three regions in the phase diagram. We predict for the red-detuned situation ($\Delta_c<0$) the formation of a spatial lattice along the pump direction accompanied with a steady state of the cavity field. By mapping the relevant degrees of freedom to a three-level Dicke-like model \cite{3Dicke1,3Dicke2,3dicke3,Dickereview}, we argue  that  a  novel  type  of  instability  phase  transition separates this regime from a blue-detuned one ($\Delta_c>0$), where we predict for small pumping strengths the formation of a dissipative spatio-temporal lattice with an oscillating cavity field. For high enough pumping strengths, the cavity field oscillations dephase, giving rise to a third region in the phase diagram. The relevant order parameter for these phases is the Fourier transform of the photon amplitude $\alpha (t)$, which is directly measurable due to the inherent leakage of photons from the cavity.  

All self-ordered phases are leading to an accelerated (super-ballistic) transport of the atoms~\cite{zheng2018anomalous} in the propagation direction of the pump field. Generally, in the presence of a lattice and driving, one would expect Bloch oscillations \cite{BlochScience} and no transport. However, the acceleration observed in our system is a result of momentum conservation in presence of cavity losses, and is proportional to the cavity field dissipation rate  $\kappa$. Our results are given by mean-field theory, which we prove to be exact in the relevant parameter ranges, and confirmed with numerical simulations and a $SU(3)$ Holstein-Primakoff transformation.


\emph{Driven Bose-Einstein condensate (BEC) in a dissipative cavity—} We consider a density matrix $\rho$ in the Lindblad master equation framework. By going into the  frame rotating at the atomic transition frequency and eliminating the excited level of the atoms, we start with the time-dependent master equation
\begin{equation}
    \begin{aligned}
    \frac{d}{dt}\rho(t)&=\hat{\mathcal{L}}\rho(t)\\
    :&=-i[H,\rho(t)]+\kappa[2a\rho(t) a^\dag -\{a^\dag a,\rho(t)\}]
    \end{aligned}
\label{MasterEqn}
\end{equation}
with
\begin{equation}
    H=-\hbar \Delta_c a^\dag a+T_{\rm{atom}}+\eta_p(a^\dag O+ aO^\dag)\,,
\label{Hamiltonian}
\end{equation}
where \(a\) (\(a^\dag\)) is the annihilation (creation) operator of a cavity photon and \(T_{\rm{atom}}=\sum_q \frac{(\hbar q)^2}{2m}c_{q}^\dag c_{q}\) is the kinetic energy of the atoms. The operator \(
    O=\sum_q c_{q+k}^\dag c_{q}=\int dx \Psi^\dag(x) e^{ikx}\Psi (x)
\) characterizes the spatial overlap of the atomic field with the electric field of the pump laser with wave vector \(k=2\pi/\lambda\), where $\lambda$ is the wave length of the pump field. Here, \(\Psi(x)\) and \(c_q\) are the annihilation operators of an atom at position \(x\) and with momentum \(q\), which are connected via the Fourier transform \(\Psi(x)=\sum_{q}e^{iq\cdot x}c_q/\sqrt{L}\). The expression \(a^\dag O\) captures the collective absorption of one recoil momentum \(\hbar k\) of the atoms by scattering one photon from the pump field into the resonator, while \(a O^\dag\) describes the inverse process annihilating one cavity photon. The pumping strength \(\eta_p\) gives the rate at which these processes take place. Since also cavity dissipation at rate \(\kappa\) leads to the annihilation of cavity photons, there is a net momentum transfer onto the atomic system in the direction of the propagation of the pumping field. Note that we neglect in our one dimensional description any momentum transfer in the direction of the cavity.

The equations of motion for the atomic momentum and cavity photon operators are
\begin{equation}
\begin{aligned}
i\partial_t c_q=\frac{\hbar q^2}{2m}c_q+\eta_p( a^\dag c_{q-k}+ac_{q+k}),\\
i\partial_t a=-(\Delta_c+i\kappa) a+\eta_p \sum_qc_{q+k}^\dag c_q.
\end{aligned}
\label{EoM_kspace}
\end{equation}
We move to a mean field description by replacing the operators with complex fields amplitudes,  \(\langle a \rangle \to \alpha\) and \(\quad \langle \Psi (r) \rangle \to \sqrt{N}\phi (r),\)  and neglecting quantum fluctuations, 
\(
    \langle a\Psi^\dag (r)\Psi (r)\rangle \to N \alpha \phi^*(r) \phi (r)
\), $N$ is the number of atoms. Moreover, since the cavity field dynamics  (\(\kappa \sim\)~MHz) is much faster than the time evolution of the atomic field (\(\omega_r \sim\)~kHz), we can adiabatically eliminate the photon mode,
\(
    \langle a \rangle=\frac{N\eta_p}{\Delta_c+i\kappa}\langle O\rangle.
\)
The main results in this article are derived under this mean field approximation, whose validity can be proved with the atom-only master equation~\cite{SM}.


The atomic system scatters photons from the running wave pump field into the optical resonator, which accordingly becomes populated with a coherent field. The interference between this field and the running wave pump field gives rise to a $\lambda$-periodic optical potential in which the atomic cloud orders in a self-consistent fashion~\cite{mivehvar2021cavity}. We choose the pump laser frequency to be larger than the atomic resonance frequency, such that atoms are expelled from regions with high intensities. The density modulated atomic cloud nevertheless scatters efficiently into the resonator since the two fields interfere destructively at the position of the maxima of the atomic density, creating local potential minima for the atoms. This is true for a red detuning  $\Delta_c<0$ between pump field and cavity resonance, where the cavity field follows the drive field in phase, and a steady state intra-cavity field can build up. In the blue detuned case $\Delta_c>0$, however, an additional $\pi$ phase shift between the driving and the scattered fields is introduced which gives rise to a non-stationary evolution. We separate the following discussion into the red and blue detuned cases for $\Delta_c$.

We study the system numerically in the coordinate basis (following adiabatic elimination) and analytically in the momentum basis both with and without adiabatic elimination of the cavity field. We use both approaches due to the fact that any finite number of modes cannot fully capture the long-time dynamics, as we will see. Physically, this is a manifestation of the persistent acceleration. First, we study the red detuned regime where a non-linearity causes spatial modulation for any finite pumping.

\begin{figure}[ht]
\centering
\includegraphics[]{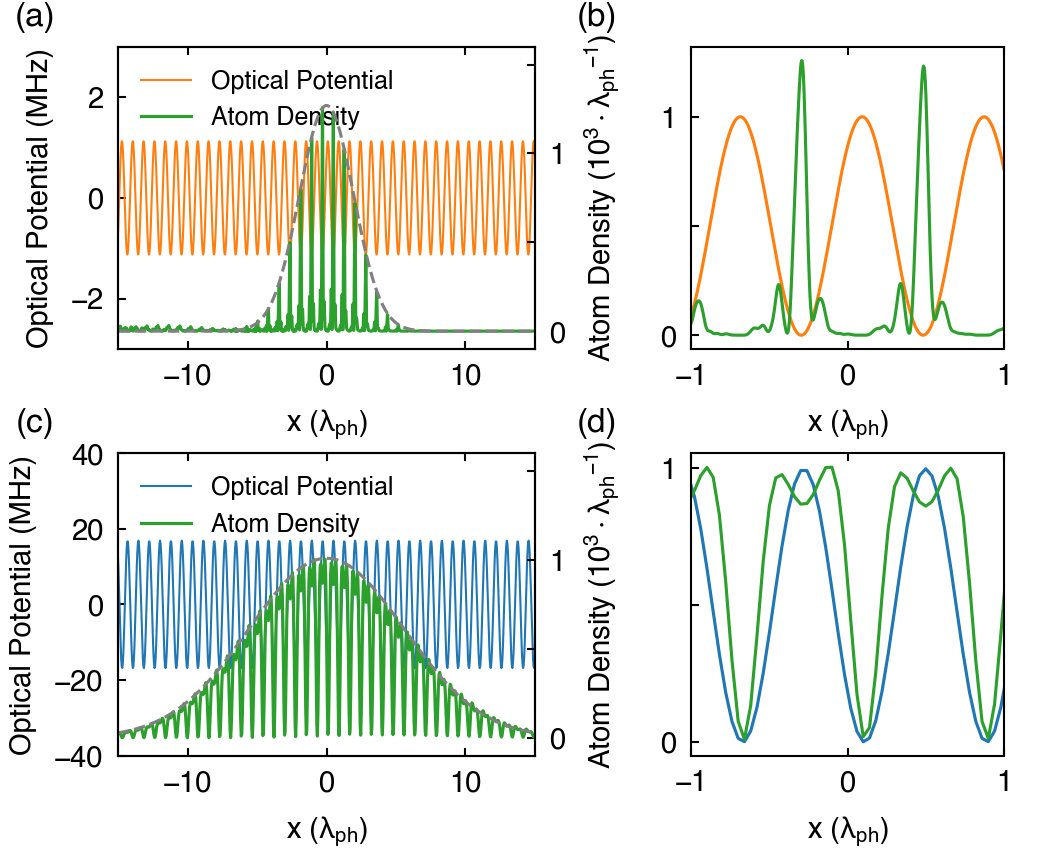}
\caption{(a) (b) In the red detuned case $\Delta_c<0$, the atomic wave function (green) localizes close to the minima of the optical potential (red). Non-zero cavity dissipation leads to a relative shift between the two which induces a constant force accelerating the atoms in the positive direction. (c) (d) In the blue detuned case $\Delta_c>0$, the additional phase shift by the cavity leads to an instable situation where the atomic wave function localizes at the maxima of the optical potential. A coupling to higher momentum states is induced which make the system form a dissipative spatio-temporal lattice. Shown is a snapshot of the evolution where two momentum states are simultaneously populated and the optical potential becomes maximal.  }
\label{fig:SpatialPotentials}
\end{figure}

\begin{figure*}[ht]
\includegraphics[]{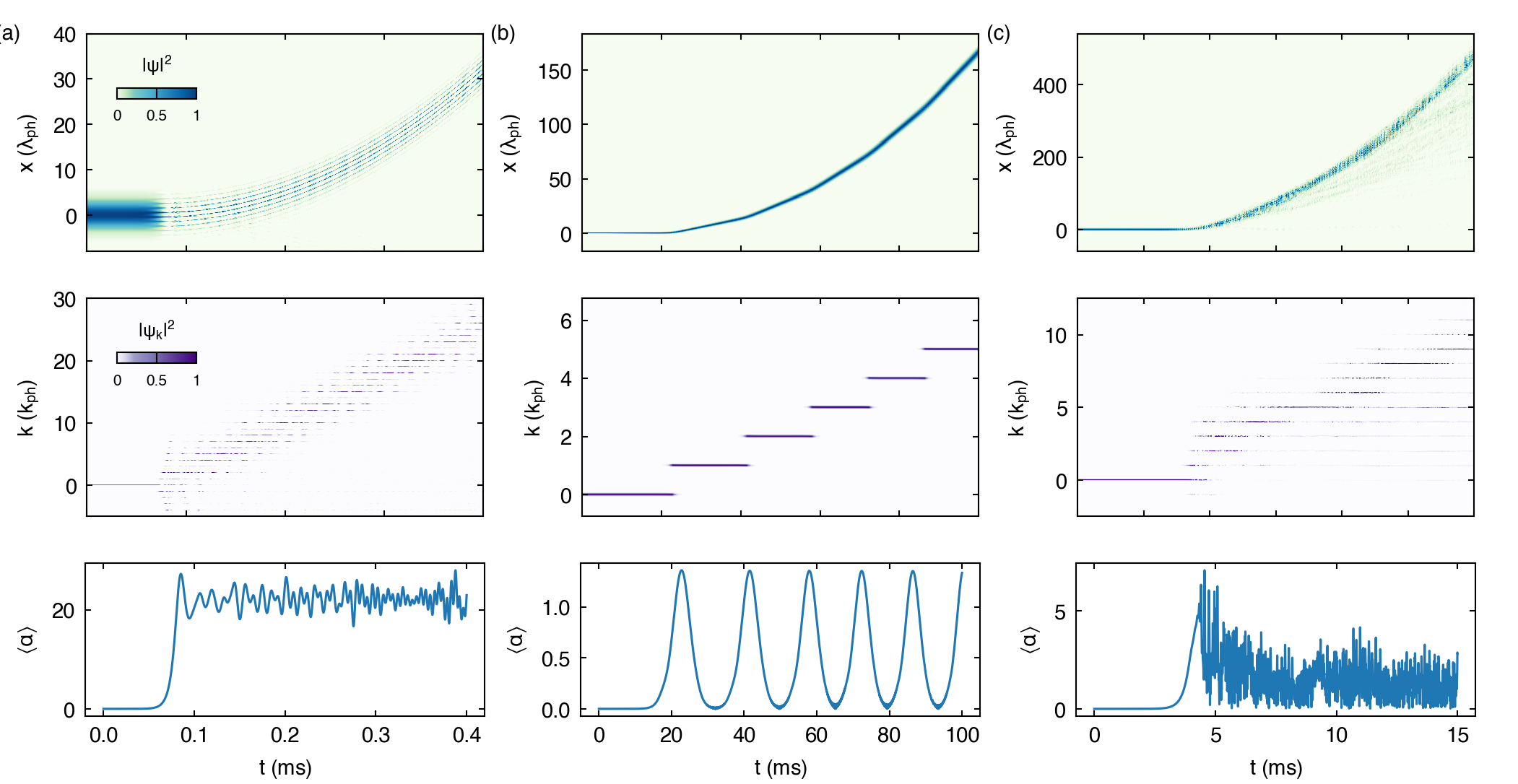}
\caption{\label{fig:Dynamics} Time evolution of the atomic wave packet in real space (upper row) and in momentum space (middle row) together with the evolution of the mean cavity field amplitude (lowest row) in the three different regimes. (a) Red detuned regime $\Delta_c<0$. The atoms are continuously accelerated and the cavity is in a steady state. (b) Blue detuned regime $\Delta_c>0$ for small pumping strengths $\eta_p$. The atomic state is consecutively climbing up the momentum ladder in a step like fashion, driven by a cavity field that periodically switches on and off. (c) Blue detuned regime $\Delta_c>0$ for large pumping strengths $\eta_p$. The strong driving strength induces simultaneous coupling to multiple momentum states which leads to a dephasing of the cavity field.}
\end{figure*}

\emph{Spatial lattice in the red detuned regime—}
We study the system in the regime $\Delta_c<0$ using as ansatz a Bloch wave function modulated by an envelope function (such as a Gaussian wave packet): \( \phi(x)=f(x-x_0)e^{iqx}u_q(x-x_0)  \). Here, \(x_0\) is the center of the wave packet, \(q\) is the quasi-momentum, and \(u_q(x)\) is an even periodic function that satisfies \(u_q(x)=u_q(x+2\pi/k)\). The slow-varying envelope function \(f(x)\) satisfies \(|(\partial_x f(x))/f(x)|\ll k\), and we assume that the maximum of \(u_q(x)\) is located at \(x=0\). We adiabatically eliminate the fast evolving cavity field and find the value $\alpha_0$ for the steady state cavity field amplitude:
\begin{equation}
    \alpha_0\approx \frac{N\eta_p}{\Delta_c+i\kappa}\ e^{ikx_0}\cdot \frac{k}{2\pi} \int_{-\frac{\pi}{k}}^{\frac{\pi}{k}}\cos(kx)|u_q(x)|^2dx\,.
\label{CavityField}
\end{equation}
Since the integral is real, only the term \((\Delta_c+i\kappa)^{-1}\) contributes to the phase of \(\alpha_0\), such that the resulting optical potential is  $V=\eta_p N|\alpha_0|\cos[k(x-x_0)-\arg(\frac{1}{\Delta_c+i\kappa})]$. For \(\kappa=0\), each Wannier component of function \(u_q(x)\) is localized at the minima of the self-consistent periodic potential, and both cavity field and the density modulated atomic cloud are stationary.

However, for non-zero cavity dissipation, a phase shift \(\delta=\tan^{-1}(\kappa/|\Delta_c|)\) is introduced, such that the maxima of the atomic wave function do not coincide with the minima of the self-generated optical potential, see Fig. \ref{fig:SpatialPotentials}a. This causes an effective force on the atomic wave packet and leads to transport in the \(+x\) direction. Neglecting the shape of the Bloch wave function, we approximate this force to~\cite{SM}
\begin{equation}
    \Bigl\langle \frac{dP}{dt} \Bigr\rangle \approx\frac{\kappa k}{2}\frac{(N\eta_p)^2}{|\Delta_c|^2+\kappa^2}.
\label{Acceleration}
\end{equation}
After building up the intra-cavity field, the periodic modulation of the atomic wave function and the self-generated potential leading to transport remain co-moving. Thus the force acting on the atomic system leads to a constant acceleration of the atomic cloud which preserves its periodic pattern. We show the time evolution of the atomic wave packet in real and in momentum space together with the cavity field amplitude in Fig.~\ref{fig:Dynamics}a.
\\

\emph{Dissipative spatio-temporal lattice in the blue detuned regime---}
For blue detuning $\Delta_c>0$, the atomic Bloch wave packet in the emerging periodic potential is instable due to the additional $\pi$ phase shift of the cavity field. Each Wannier component of the atomic wave function now tries to localize at the maxima of the optical potential and no stationary state is formed, hence the previous description fails. Instead, we work in the momentum space and choose all \(N\) atoms to be initially in the coherent momentum state \({\mid} q \rangle\). Only the neighboring momentum states \({\mid} q-k \rangle\) and \({\mid} q+k \rangle\) are directly coupled to the initial state \({\mid} q \rangle\). By studying the short-time dynamics of the system (see Supplementary Material), the dissipation suppresses the transition \(q \to q-k\) and only the \(q \to q+k\) transition is allowed in the blue-detuned regime. So, we can study the dynamics in the blue-detuned regime by truncating to two neighbouring momentum states. 

For weak pumping, the equations of motion for the states  \({\mid} q\rangle\) and \({\mid} q+k \rangle\) are
\begin{equation}
    i\partial_t 
    \left(\begin{array}{c}
        \langle c_{q} \rangle  \\
        \langle c_{q+k} \rangle 
    \end{array}\right)
    =
    \left(\begin{array}{cc}
        \omega_{q} & \eta_p \alpha   \\
        \eta_p \alpha^* & \omega_{q+k}   
    \end{array}\right)
    \left(\begin{array}{c}
        \langle c_{q} \rangle   \\
        \langle c_{q+k} \rangle  
    \end{array}\right)\,, 
\label{2ModesModel}
\end{equation}
where the cavity field amplitude \(\alpha=\frac{N\eta_q}{\Delta_c+i\kappa}\langle c_{q+k}^\dag\rangle\langle c_q\rangle\) importantly depends on the product of the population of the two involved momentum states. This truncation to two momentum states fails completely for the red-detuned case even in the short-time limit (see Fig.~\ref{fig:Dynamics}). For weak coupling \(\eta_p \alpha\), the dynamics of this effective two-level system is a Rabi oscillation. However, once a "\(\pi\)-pulse" has been applied, the occupation of the \({\mid} q\rangle\) state drops to zero and the driving accordingly turns off. Now, the \({\mid} q+k \rangle\) state is macroscopically occupied and a new cycle of transition between the states \({\mid} q+k \rangle\) and \({\mid} q+2k \rangle\) begins and dissipation prevents return to lower momentum states. Figure~\ref{fig:Dynamics}b displays real space and momentum space evolutions of the system together with the cavity field amplitude. This field shows a periodic evolution. It is populated only during the transitions between neighboring momentum states and vanishes 
when one momentum state is fully occupied by all atoms. Accordingly, the system is accelerated in a step-like fashion, different from the constant acceleration for red detuning. The atomic cloud is periodically changing its shape, as we illustrate in Fig.~\ref{fig:SpatialPotentials} where the system populates simultaneously two different momentum states. Hence we call this non-stationary phase of the system a dissipative spatio-temporal lattice.

The energy spacing between two neighboring momentum states is given by the recoil energy $\hbar \omega_r$. Moving away from the weak coupling limit  \(
    \eta_q\alpha(t) \leq \frac{N\eta_p^2}{4 \mid \Delta_c+i\kappa \mid} \ll \omega_r 
\)  by either applying a strong pump $\eta_p$ or a small cavity detuning $\Delta_c$, transitions to non-neighboring momentum states can also be induced. Once these states are sufficiently occupied, additional frequency components emerge and cause a dephasing of the cavity field such that the oscillatory behavior breaks down. In this situation the dissipative spatio-temporal lattice disappears and a phase with a dephased cavity field and simultaneous occupation of multiple momentum states with increasing momenta emerges, see Fig.~\ref{fig:Dynamics}. Again, the system climbs up the momentum ladder and is accordingly accelerated. This dephasing regime cannot be captured by truncation to any finite number of modes because the incommensurate frequency (dephasing) is due to the presence of a large number of modes and we study it numerically.
\\

\begin{figure}[ht]
\centering
\includegraphics[]{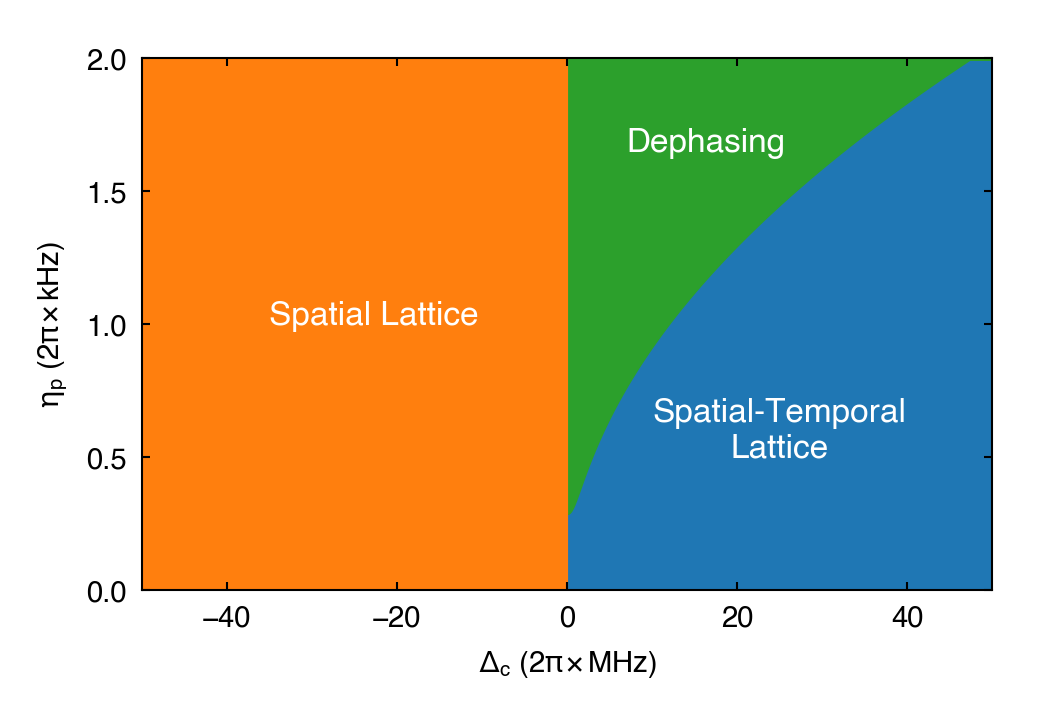}
\caption{The dynamical phase diagram of the system with atom number \(N=10^5\) and the cavity dissipation \(\kappa=2\pi\times 1\)MHz. For negative cavity detuning the spatial lattice is formed for any finite pumping strength (red regime). Under large positive cavity detuning and weak pumping, oscillation in cavity field occur (dissipative spatio-temporal lattice, blue regime). Dephasing occurs under small cavity detuning and strong pumping (dephasing, blue regime). The phase boundary is \((8\omega_r|\Delta_c+i\kappa|)/(N\eta_p^2)=3.5\) between the spatio-temporal lattice and dephasing is obtained numerically. }
\label{fig:PhaseDiagram}
\end{figure}
\emph{Phase diagram---}Finally, we map out the phase diagram and identify three different regimes. The phases are characterized by the time evolution of the intra-cavity light field which is either in a steady state, shows periodic pulsing, or has a dephased evolution. The resulting phase diagram as a function of cavity detuning $\Delta_c$ and pump strength $\eta_p$ is shown in Fig.~\ref{fig:PhaseDiagram}. We  studied the phase diagram more precisely by mapping the full model to a three-level Dicke-like model. As we show in \cite{SM}, this model features, a novel type of phase transition between two unstable regions (the red-detuned and blue-detuned regime) corresponding to the transition identified numerically. The critical line $\Delta_c=0$ separates two different regimes with different diverging modes for non-zero $\kappa$ (in contrast to standard phase transitions between an unstable and stable region \cite{Dickereview}). For $\kappa=0$ the model reduces to a standard two-level Dicke phase transition, i.e. with an ordinary superradiant phase transition, further confirming that our spatio-temporal order is indeed dissipation induced. However, this three-mode approximation is not sufficient to capture the full long-time dynamics leading to the accelerated transport in all regimes. In particular, it cannot capture the dephasing to spatio-temporal lattice transition, which occurs due to interference (dephasing) of a very large number of momentum modes (levels) with incommensurate frequencies.
\\


\emph{Conclusion--}
Our work opens various theoretical and experimental questions and possibilities for studying novel quantum optical phases far-from-equilibrium. Although our approach is exact in the weakly-interacting limit, which is perfectly valid for the system studied, an interesting theoretical question is if the persistent oscillations in the dissipative spatio-temporal lattice survive interactions between the atoms. This, along with quantum effects, could be studied in the strong dissipation limit analogously to \cite{PhysRevLett.123.260401} using the theory of dynamical symmetries \cite{Buca_2019}. This could indicate formation of non-trivial entanglement in the system. Interactions between the atoms could similarly induce synchronization inside the lattice along the proposals in \cite{Buca_2019,quantumsynch,buca2021algebraic}. The presence of both super-ballistic transport~\cite{zheng2018anomalous} and persistent oscillations is reminiscent of persistent oscillations in the XXZ spin chain that happens only in the quantum phase with ballistic transport \cite{Marko1}. 

In the system we studied, the atoms are not confined by any lattice and trap. If we add a static lattice on the atoms, the competition between optical pumping and lattice confinement may lead to exotic dynamics of atoms such as Bloch oscillation~\cite{battesti2004bloch,dahan1996bloch} and anomalous diffusion~\cite{zheng2018anomalous,ljubotina2017spin,PhysRevLett.121.230602}. 

This work also opens up a possibility to design a cavity-assisted system for multiphoton Bragg diffraction in atom interferometers~\cite{cronin2009optics,chiow2011102}, since the BEC can achieve a high momentum state under a very weak pumping while the evolution is coherent. With two counter-propagating beams modulated by finely optimized pulses~\cite{hohenester2007optimal}, it may be possible to coherently prepare the BEC into a superposition of an arbitrary set of momentum states to perform atom interference.

\begin{acknowledgements}
\emph{Acknowledgment} We thank V. Juki\'{c} Bu\v{c}a for help with Fig.~1, and Alexander Baumgärtner and Simon Hertlein for useful discussions. This work is supported by the Cluster of Excellence 'CUI: Advanced Imaging of Matter' of the Deutsche Forschungsgemeinschaft (DFG) - EXC 2056 - project ID 390715994, EPSRC programme grants EP/P009565/1, EP/P01058X/1, EPSRC National Quantum Technology Hub in Networked Quantum Information Technology (EP/M013243/1), the SNF projects 182650, 175329 (NAQUAS QuantERA), IZBRZ2 186312, and NCCR QSIT.
\end{acknowledgements}


\bibliographystyle{apsrev4-2}
\bibliography{apssamp}

\begin{thebibliography}{64}%
\makeatletter
\providecommand \@ifxundefined [1]{%
 \@ifx{#1\undefined}
}%
\providecommand \@ifnum [1]{%
 \ifnum #1\expandafter \@firstoftwo
 \else \expandafter \@secondoftwo
 \fi
}%
\providecommand \@ifx [1]{%
 \ifx #1\expandafter \@firstoftwo
 \else \expandafter \@secondoftwo
 \fi
}%
\providecommand \natexlab [1]{#1}%
\providecommand \enquote  [1]{``#1''}%
\providecommand \bibnamefont  [1]{#1}%
\providecommand \bibfnamefont [1]{#1}%
\providecommand \citenamefont [1]{#1}%
\providecommand \href@noop [0]{\@secondoftwo}%
\providecommand \href [0]{\begingroup \@sanitize@url \@href}%
\providecommand \@href[1]{\@@startlink{#1}\@@href}%
\providecommand \@@href[1]{\endgroup#1\@@endlink}%
\providecommand \@sanitize@url [0]{\catcode `\\12\catcode `\$12\catcode
  `\&12\catcode `\#12\catcode `\^12\catcode `\_12\catcode `\%12\relax}%
\providecommand \@@startlink[1]{}%
\providecommand \@@endlink[0]{}%
\providecommand \url  [0]{\begingroup\@sanitize@url \@url }%
\providecommand \@url [1]{\endgroup\@href {#1}{\urlprefix }}%
\providecommand \urlprefix  [0]{URL }%
\providecommand \Eprint [0]{\href }%
\providecommand \doibase [0]{https://doi.org/}%
\providecommand \selectlanguage [0]{\@gobble}%
\providecommand \bibinfo  [0]{\@secondoftwo}%
\providecommand \bibfield  [0]{\@secondoftwo}%
\providecommand \translation [1]{[#1]}%
\providecommand \BibitemOpen [0]{}%
\providecommand \bibitemStop [0]{}%
\providecommand \bibitemNoStop [0]{.\EOS\space}%
\providecommand \EOS [0]{\spacefactor3000\relax}%
\providecommand \BibitemShut  [1]{\csname bibitem#1\endcsname}%
\let\auto@bib@innerbib\@empty
\bibitem [{\citenamefont {Cardy}(1996)}]{cardy_1996}%
  \BibitemOpen
  \bibfield  {author} {\bibinfo {author} {\bibfnamefont {J.}~\bibnamefont
  {Cardy}},\ }\href {https://doi.org/10.1017/CBO9781316036440} {\emph {\bibinfo
  {title} {Scaling and Renormalization in Statistical Physics}}},\ Cambridge
  Lecture Notes in Physics\ (\bibinfo  {publisher} {Cambridge University
  Press},\ \bibinfo {year} {1996})\BibitemShut {NoStop}%
\bibitem [{\citenamefont {Wilczek}(2012)}]{Wilczek}%
  \BibitemOpen
  \bibfield  {author} {\bibinfo {author} {\bibfnamefont {F.}~\bibnamefont
  {Wilczek}},\ }\href {https://doi.org/10.1103/PhysRevLett.109.160401}
  {\bibfield  {journal} {\bibinfo  {journal} {Phys. Rev. Lett.}\ }\textbf
  {\bibinfo {volume} {109}},\ \bibinfo {pages} {160401} (\bibinfo {year}
  {2012})}\BibitemShut {NoStop}%
\bibitem [{\citenamefont {Barberena}\ \emph {et~al.}(2019)\citenamefont
  {Barberena}, \citenamefont {Lewis-Swan}, \citenamefont {Thompson},\ and\
  \citenamefont {Rey}}]{Dissipative1}%
  \BibitemOpen
  \bibfield  {author} {\bibinfo {author} {\bibfnamefont {D.}~\bibnamefont
  {Barberena}}, \bibinfo {author} {\bibfnamefont {R.~J.}\ \bibnamefont
  {Lewis-Swan}}, \bibinfo {author} {\bibfnamefont {J.~K.}\ \bibnamefont
  {Thompson}},\ and\ \bibinfo {author} {\bibfnamefont {A.~M.}\ \bibnamefont
  {Rey}},\ }\href {https://doi.org/10.1103/PhysRevA.99.053411} {\bibfield
  {journal} {\bibinfo  {journal} {Phys. Rev. A}\ }\textbf {\bibinfo {volume}
  {99}},\ \bibinfo {pages} {053411} (\bibinfo {year} {2019})}\BibitemShut
  {NoStop}%
\bibitem [{\citenamefont {Scarlatella}\ \emph {et~al.}(2019)\citenamefont
  {Scarlatella}, \citenamefont {Fazio},\ and\ \citenamefont
  {Schir\'o}}]{Orazio}%
  \BibitemOpen
  \bibfield  {author} {\bibinfo {author} {\bibfnamefont {O.}~\bibnamefont
  {Scarlatella}}, \bibinfo {author} {\bibfnamefont {R.}~\bibnamefont {Fazio}},\
  and\ \bibinfo {author} {\bibfnamefont {M.}~\bibnamefont {Schir\'o}},\ }\href
  {https://doi.org/10.1103/PhysRevB.99.064511} {\bibfield  {journal} {\bibinfo
  {journal} {Phys. Rev. B}\ }\textbf {\bibinfo {volume} {99}},\ \bibinfo
  {pages} {064511} (\bibinfo {year} {2019})}\BibitemShut {NoStop}%
\bibitem [{\citenamefont {Chiacchio}\ and\ \citenamefont
  {Nunnenkamp}(2019)}]{Nunnenkamp}%
  \BibitemOpen
  \bibfield  {author} {\bibinfo {author} {\bibfnamefont {E.~I.~R.}\
  \bibnamefont {Chiacchio}}\ and\ \bibinfo {author} {\bibfnamefont
  {A.}~\bibnamefont {Nunnenkamp}},\ }\href
  {https://doi.org/10.1103/PhysRevLett.122.193605} {\bibfield  {journal}
  {\bibinfo  {journal} {Phys. Rev. Lett.}\ }\textbf {\bibinfo {volume} {122}},\
  \bibinfo {pages} {193605} (\bibinfo {year} {2019})}\BibitemShut {NoStop}%
\bibitem [{\citenamefont {Seibold}\ \emph {et~al.}(2020)\citenamefont
  {Seibold}, \citenamefont {Rota},\ and\ \citenamefont {Savona}}]{Seibold}%
  \BibitemOpen
  \bibfield  {author} {\bibinfo {author} {\bibfnamefont {K.}~\bibnamefont
  {Seibold}}, \bibinfo {author} {\bibfnamefont {R.}~\bibnamefont {Rota}},\ and\
  \bibinfo {author} {\bibfnamefont {V.}~\bibnamefont {Savona}},\ }\href
  {https://doi.org/10.1103/PhysRevA.101.033839} {\bibfield  {journal} {\bibinfo
   {journal} {Phys. Rev. A}\ }\textbf {\bibinfo {volume} {101}},\ \bibinfo
  {pages} {033839} (\bibinfo {year} {2020})}\BibitemShut {NoStop}%
\bibitem [{\citenamefont {Oberreiter}\ \emph {et~al.}(2021)\citenamefont
  {Oberreiter}, \citenamefont {Seifert},\ and\ \citenamefont
  {Barato}}]{Seifert}%
  \BibitemOpen
  \bibfield  {author} {\bibinfo {author} {\bibfnamefont {L.}~\bibnamefont
  {Oberreiter}}, \bibinfo {author} {\bibfnamefont {U.}~\bibnamefont
  {Seifert}},\ and\ \bibinfo {author} {\bibfnamefont {A.~C.}\ \bibnamefont
  {Barato}},\ }\href {https://doi.org/10.1103/PhysRevLett.126.020603}
  {\bibfield  {journal} {\bibinfo  {journal} {Phys. Rev. Lett.}\ }\textbf
  {\bibinfo {volume} {126}},\ \bibinfo {pages} {020603} (\bibinfo {year}
  {2021})}\BibitemShut {NoStop}%
\bibitem [{\citenamefont {Mendoza-Arenas}\ and\ \citenamefont
  {Buča}(2021)}]{JuanJose2}%
  \BibitemOpen
  \bibfield  {author} {\bibinfo {author} {\bibfnamefont {J.~J.}\ \bibnamefont
  {Mendoza-Arenas}}\ and\ \bibinfo {author} {\bibfnamefont {B.}~\bibnamefont
  {Buča}},\ }\href@noop {} {\bibinfo {title} {Self-induced entanglement
  resonance in a disordered bose-fermi mixture}} (\bibinfo {year} {2021}),\
  \Eprint {https://arxiv.org/abs/2106.06277} {arXiv:2106.06277
  [cond-mat.quant-gas]} \BibitemShut {NoStop}%
\bibitem [{\citenamefont {Minganti}\ \emph {et~al.}(2020)\citenamefont
  {Minganti}, \citenamefont {Arkhipov}, \citenamefont {Miranowicz},\ and\
  \citenamefont {Nori}}]{Fabrizio}%
  \BibitemOpen
  \bibfield  {author} {\bibinfo {author} {\bibfnamefont {F.}~\bibnamefont
  {Minganti}}, \bibinfo {author} {\bibfnamefont {I.~I.}\ \bibnamefont
  {Arkhipov}}, \bibinfo {author} {\bibfnamefont {A.}~\bibnamefont
  {Miranowicz}},\ and\ \bibinfo {author} {\bibfnamefont {F.}~\bibnamefont
  {Nori}},\ }\href@noop {} {\bibinfo {title} {Correspondence between
  dissipative phase transitions of light and time crystals}} (\bibinfo {year}
  {2020}),\ \Eprint {https://arxiv.org/abs/2008.08075} {arXiv:2008.08075
  [quant-ph]} \BibitemShut {NoStop}%
\bibitem [{\citenamefont {Carollo}\ and\ \citenamefont
  {Lesanovsky}(2021)}]{Lesanovsky}%
  \BibitemOpen
  \bibfield  {author} {\bibinfo {author} {\bibfnamefont {F.}~\bibnamefont
  {Carollo}}\ and\ \bibinfo {author} {\bibfnamefont {I.}~\bibnamefont
  {Lesanovsky}},\ }\href@noop {} {\bibinfo {title} {Exact solution of a
  boundary time-crystal phase transition: time-translation symmetry breaking
  and non-markovian dynamics of correlations}} (\bibinfo {year} {2021}),\
  \Eprint {https://arxiv.org/abs/2110.00030} {arXiv:2110.00030
  [cond-mat.stat-mech]} \BibitemShut {NoStop}%
\bibitem [{\citenamefont {Seetharam}\ \emph {et~al.}(2021)\citenamefont
  {Seetharam}, \citenamefont {Lerose}, \citenamefont {Fazio},\ and\
  \citenamefont {Marino}}]{Jamir3}%
  \BibitemOpen
  \bibfield  {author} {\bibinfo {author} {\bibfnamefont {K.}~\bibnamefont
  {Seetharam}}, \bibinfo {author} {\bibfnamefont {A.}~\bibnamefont {Lerose}},
  \bibinfo {author} {\bibfnamefont {R.}~\bibnamefont {Fazio}},\ and\ \bibinfo
  {author} {\bibfnamefont {J.}~\bibnamefont {Marino}},\ }\href@noop {}
  {\bibinfo {title} {Correlation engineering via non-local dissipation}}
  (\bibinfo {year} {2021}),\ \Eprint {https://arxiv.org/abs/2101.06445}
  {arXiv:2101.06445 [cond-mat.quant-gas]} \BibitemShut {NoStop}%
\bibitem [{\citenamefont {Chen}\ and\ \citenamefont
  {Navarrete-Benlloch}(2021)}]{chen2021collectively}%
  \BibitemOpen
  \bibfield  {author} {\bibinfo {author} {\bibfnamefont {Y.}~\bibnamefont
  {Chen}}\ and\ \bibinfo {author} {\bibfnamefont {C.}~\bibnamefont
  {Navarrete-Benlloch}},\ }\href@noop {} {\bibinfo {title} {Collectively
  pair-driven-dissipative bosonic arrays: exotic and self-oscillatory
  condensates}} (\bibinfo {year} {2021}),\ \Eprint
  {https://arxiv.org/abs/2111.07326} {arXiv:2111.07326 [cond-mat.quant-gas]}
  \BibitemShut {NoStop}%
\bibitem [{\citenamefont {Seibold}\ \emph {et~al.}(2021)\citenamefont
  {Seibold}, \citenamefont {Rota}, \citenamefont {Minganti},\ and\
  \citenamefont {Savona}}]{seibold2021quantum}%
  \BibitemOpen
  \bibfield  {author} {\bibinfo {author} {\bibfnamefont {K.}~\bibnamefont
  {Seibold}}, \bibinfo {author} {\bibfnamefont {R.}~\bibnamefont {Rota}},
  \bibinfo {author} {\bibfnamefont {F.}~\bibnamefont {Minganti}},\ and\
  \bibinfo {author} {\bibfnamefont {V.}~\bibnamefont {Savona}},\ }\href@noop {}
  {\bibinfo {title} {Quantum dynamics of dissipative kerr solitons}} (\bibinfo
  {year} {2021}),\ \Eprint {https://arxiv.org/abs/2112.00611} {arXiv:2112.00611
  [quant-ph]} \BibitemShut {NoStop}%
\bibitem [{\citenamefont {Buča}\ \emph {et~al.}(2019)\citenamefont {Buča},
  \citenamefont {Tindall},\ and\ \citenamefont {Jaksch}}]{Buca_2019}%
  \BibitemOpen
  \bibfield  {author} {\bibinfo {author} {\bibfnamefont {B.}~\bibnamefont
  {Buča}}, \bibinfo {author} {\bibfnamefont {J.}~\bibnamefont {Tindall}},\
  and\ \bibinfo {author} {\bibfnamefont {D.}~\bibnamefont {Jaksch}},\ }\href
  {https://doi.org/10.1038/s41467-019-09757-y} {\bibfield  {journal} {\bibinfo
  {journal} {Nature Communications}\ }\textbf {\bibinfo {volume} {10}},\
  \bibinfo {pages} {1730} (\bibinfo {year} {2019})}\BibitemShut {NoStop}%
\bibitem [{\citenamefont {Booker}\ \emph {et~al.}(2020)\citenamefont {Booker},
  \citenamefont {Buča},\ and\ \citenamefont {Jaksch}}]{Booker_2020}%
  \BibitemOpen
  \bibfield  {author} {\bibinfo {author} {\bibfnamefont {C.}~\bibnamefont
  {Booker}}, \bibinfo {author} {\bibfnamefont {B.}~\bibnamefont {Buča}},\ and\
  \bibinfo {author} {\bibfnamefont {D.}~\bibnamefont {Jaksch}},\ }\bibfield
  {journal} {\bibinfo  {journal} {New Journal of Physics}\ }\href
  {https://doi.org/10.1088/1367-2630/ababc4} {10.1088/1367-2630/ababc4}
  (\bibinfo {year} {2020})\BibitemShut {NoStop}%
\bibitem [{\citenamefont {Iemini}\ \emph {et~al.}(2018)\citenamefont {Iemini},
  \citenamefont {Russomanno}, \citenamefont {Keeling}, \citenamefont
  {Schir\`o}, \citenamefont {Dalmonte},\ and\ \citenamefont {Fazio}}]{Fazio}%
  \BibitemOpen
  \bibfield  {author} {\bibinfo {author} {\bibfnamefont {F.}~\bibnamefont
  {Iemini}}, \bibinfo {author} {\bibfnamefont {A.}~\bibnamefont {Russomanno}},
  \bibinfo {author} {\bibfnamefont {J.}~\bibnamefont {Keeling}}, \bibinfo
  {author} {\bibfnamefont {M.}~\bibnamefont {Schir\`o}}, \bibinfo {author}
  {\bibfnamefont {M.}~\bibnamefont {Dalmonte}},\ and\ \bibinfo {author}
  {\bibfnamefont {R.}~\bibnamefont {Fazio}},\ }\href
  {https://doi.org/10.1103/PhysRevLett.121.035301} {\bibfield  {journal}
  {\bibinfo  {journal} {Phys. Rev. Lett.}\ }\textbf {\bibinfo {volume} {121}},\
  \bibinfo {pages} {035301} (\bibinfo {year} {2018})}\BibitemShut {NoStop}%
\bibitem [{\citenamefont {Sarkar}\ and\ \citenamefont
  {Dubi}(2021)}]{sarkar2021signatures}%
  \BibitemOpen
  \bibfield  {author} {\bibinfo {author} {\bibfnamefont {S.}~\bibnamefont
  {Sarkar}}\ and\ \bibinfo {author} {\bibfnamefont {Y.}~\bibnamefont {Dubi}},\
  }\href@noop {} {\bibinfo {title} {Signatures of discrete time-crystallinity
  in transport through quantum dot arrays}} (\bibinfo {year} {2021}),\ \Eprint
  {https://arxiv.org/abs/2107.04214} {arXiv:2107.04214 [cond-mat.mes-hall]}
  \BibitemShut {NoStop}%
\bibitem [{\citenamefont {Ritsch}\ \emph {et~al.}(2013)\citenamefont {Ritsch},
  \citenamefont {Domokos}, \citenamefont {Brennecke},\ and\ \citenamefont
  {Esslinger}}]{Cavity2}%
  \BibitemOpen
  \bibfield  {author} {\bibinfo {author} {\bibfnamefont {H.}~\bibnamefont
  {Ritsch}}, \bibinfo {author} {\bibfnamefont {P.}~\bibnamefont {Domokos}},
  \bibinfo {author} {\bibfnamefont {F.}~\bibnamefont {Brennecke}},\ and\
  \bibinfo {author} {\bibfnamefont {T.}~\bibnamefont {Esslinger}},\ }\href
  {https://doi.org/10.1103/RevModPhys.85.553} {\bibfield  {journal} {\bibinfo
  {journal} {Rev. Mod. Phys.}\ }\textbf {\bibinfo {volume} {85}},\ \bibinfo
  {pages} {553} (\bibinfo {year} {2013})}\BibitemShut {NoStop}%
\bibitem [{\citenamefont {Mivehvar}\ \emph
  {et~al.}(2021{\natexlab{a}})\citenamefont {Mivehvar}, \citenamefont {Piazza},
  \citenamefont {Donner},\ and\ \citenamefont {Ritsch}}]{cavityreview}%
  \BibitemOpen
  \bibfield  {author} {\bibinfo {author} {\bibfnamefont {F.}~\bibnamefont
  {Mivehvar}}, \bibinfo {author} {\bibfnamefont {F.}~\bibnamefont {Piazza}},
  \bibinfo {author} {\bibfnamefont {T.}~\bibnamefont {Donner}},\ and\ \bibinfo
  {author} {\bibfnamefont {H.}~\bibnamefont {Ritsch}},\ }\href@noop {}
  {\bibinfo {title} {Cavity qed with quantum gases: New paradigms in many-body
  physics}} (\bibinfo {year} {2021}{\natexlab{a}}),\ \Eprint
  {https://arxiv.org/abs/2102.04473} {arXiv:2102.04473 [cond-mat.quant-gas]}
  \BibitemShut {NoStop}%
\bibitem [{\citenamefont {Lin}\ \emph {et~al.}(2021)\citenamefont {Lin},
  \citenamefont {Rosa-Medina}, \citenamefont {Ferri}, \citenamefont {Finger},
  \citenamefont {Kroeger}, \citenamefont {Donner}, \citenamefont {Esslinger},\
  and\ \citenamefont {Chitra}}]{Cavity4}%
  \BibitemOpen
  \bibfield  {author} {\bibinfo {author} {\bibfnamefont {R.}~\bibnamefont
  {Lin}}, \bibinfo {author} {\bibfnamefont {R.}~\bibnamefont {Rosa-Medina}},
  \bibinfo {author} {\bibfnamefont {F.}~\bibnamefont {Ferri}}, \bibinfo
  {author} {\bibfnamefont {F.}~\bibnamefont {Finger}}, \bibinfo {author}
  {\bibfnamefont {K.}~\bibnamefont {Kroeger}}, \bibinfo {author} {\bibfnamefont
  {T.}~\bibnamefont {Donner}}, \bibinfo {author} {\bibfnamefont
  {T.}~\bibnamefont {Esslinger}},\ and\ \bibinfo {author} {\bibfnamefont
  {R.}~\bibnamefont {Chitra}},\ }\href@noop {} {\bibinfo {title}
  {Dissipation-engineered family of nearly dark states in many-body cavity-atom
  systems}} (\bibinfo {year} {2021}),\ \Eprint
  {https://arxiv.org/abs/2109.00422} {arXiv:2109.00422 [cond-mat.quant-gas]}
  \BibitemShut {NoStop}%
\bibitem [{\citenamefont {Piazza}\ and\ \citenamefont
  {Ritsch}(2015)}]{Piazza1}%
  \BibitemOpen
  \bibfield  {author} {\bibinfo {author} {\bibfnamefont {F.}~\bibnamefont
  {Piazza}}\ and\ \bibinfo {author} {\bibfnamefont {H.}~\bibnamefont
  {Ritsch}},\ }\href {https://doi.org/10.1103/PhysRevLett.115.163601}
  {\bibfield  {journal} {\bibinfo  {journal} {Phys. Rev. Lett.}\ }\textbf
  {\bibinfo {volume} {115}},\ \bibinfo {pages} {163601} (\bibinfo {year}
  {2015})}\BibitemShut {NoStop}%
\bibitem [{\citenamefont {Alaeian}\ \emph {et~al.}(2021)\citenamefont
  {Alaeian}, \citenamefont {Soriente}, \citenamefont {Najafi},\ and\
  \citenamefont {Yelin}}]{Hadiseh}%
  \BibitemOpen
  \bibfield  {author} {\bibinfo {author} {\bibfnamefont {H.}~\bibnamefont
  {Alaeian}}, \bibinfo {author} {\bibfnamefont {M.}~\bibnamefont {Soriente}},
  \bibinfo {author} {\bibfnamefont {K.}~\bibnamefont {Najafi}},\ and\ \bibinfo
  {author} {\bibfnamefont {S.~F.}\ \bibnamefont {Yelin}},\ }\href@noop {}
  {\bibinfo {title} {Noise-resilient phase transitions and limit-cycles in
  coupled kerr oscillators}} (\bibinfo {year} {2021}),\ \Eprint
  {https://arxiv.org/abs/2106.04045} {arXiv:2106.04045 [quant-ph]} \BibitemShut
  {NoStop}%
\bibitem [{\citenamefont {Keeling}\ \emph {et~al.}(2010)\citenamefont
  {Keeling}, \citenamefont {Bhaseen},\ and\ \citenamefont {Simons}}]{Keeling}%
  \BibitemOpen
  \bibfield  {author} {\bibinfo {author} {\bibfnamefont {J.}~\bibnamefont
  {Keeling}}, \bibinfo {author} {\bibfnamefont {M.~J.}\ \bibnamefont
  {Bhaseen}},\ and\ \bibinfo {author} {\bibfnamefont {B.~D.}\ \bibnamefont
  {Simons}},\ }\href {https://doi.org/10.1103/PhysRevLett.105.043001}
  {\bibfield  {journal} {\bibinfo  {journal} {Phys. Rev. Lett.}\ }\textbf
  {\bibinfo {volume} {105}},\ \bibinfo {pages} {043001} (\bibinfo {year}
  {2010})}\BibitemShut {NoStop}%
\bibitem [{\citenamefont {Bhaseen}\ \emph {et~al.}(2012)\citenamefont
  {Bhaseen}, \citenamefont {Mayoh}, \citenamefont {Simons},\ and\ \citenamefont
  {Keeling}}]{Keeling2}%
  \BibitemOpen
  \bibfield  {author} {\bibinfo {author} {\bibfnamefont {M.~J.}\ \bibnamefont
  {Bhaseen}}, \bibinfo {author} {\bibfnamefont {J.}~\bibnamefont {Mayoh}},
  \bibinfo {author} {\bibfnamefont {B.~D.}\ \bibnamefont {Simons}},\ and\
  \bibinfo {author} {\bibfnamefont {J.}~\bibnamefont {Keeling}},\ }\href
  {https://doi.org/10.1103/PhysRevA.85.013817} {\bibfield  {journal} {\bibinfo
  {journal} {Phys. Rev. A}\ }\textbf {\bibinfo {volume} {85}},\ \bibinfo
  {pages} {013817} (\bibinfo {year} {2012})}\BibitemShut {NoStop}%
\bibitem [{\citenamefont {Baumann}\ \emph
  {et~al.}(2010{\natexlab{a}})\citenamefont {Baumann}, \citenamefont {Guerlin},
  \citenamefont {Brennecke},\ and\ \citenamefont {Esslinger}}]{Cavity1}%
  \BibitemOpen
  \bibfield  {author} {\bibinfo {author} {\bibfnamefont {K.}~\bibnamefont
  {Baumann}}, \bibinfo {author} {\bibfnamefont {C.}~\bibnamefont {Guerlin}},
  \bibinfo {author} {\bibfnamefont {F.}~\bibnamefont {Brennecke}},\ and\
  \bibinfo {author} {\bibfnamefont {T.}~\bibnamefont {Esslinger}},\ }\href
  {https://doi.org/10.1038/nature09009} {\bibfield  {journal} {\bibinfo
  {journal} {Nature}\ }\textbf {\bibinfo {volume} {464}},\ \bibinfo {pages}
  {1301–1306} (\bibinfo {year} {2010}{\natexlab{a}})}\BibitemShut {NoStop}%
\bibitem [{\citenamefont {Bu\ifmmode~\check{c}\else \v{c}\fi{}a}\ and\
  \citenamefont {Jaksch}(2019)}]{PhysRevLett.123.260401}%
  \BibitemOpen
  \bibfield  {author} {\bibinfo {author} {\bibfnamefont {B.}~\bibnamefont
  {Bu\ifmmode~\check{c}\else \v{c}\fi{}a}}\ and\ \bibinfo {author}
  {\bibfnamefont {D.}~\bibnamefont {Jaksch}},\ }\href
  {https://doi.org/10.1103/PhysRevLett.123.260401} {\bibfield  {journal}
  {\bibinfo  {journal} {Phys. Rev. Lett.}\ }\textbf {\bibinfo {volume} {123}},\
  \bibinfo {pages} {260401} (\bibinfo {year} {2019})}\BibitemShut {NoStop}%
\bibitem [{\citenamefont {Lled\'o}\ \emph {et~al.}(2019)\citenamefont
  {Lled\'o}, \citenamefont {Mavrogordatos},\ and\ \citenamefont
  {Szyma\ifmmode~\acute{n}\else \'{n}\fi{}ska}}]{Lledo1}%
  \BibitemOpen
  \bibfield  {author} {\bibinfo {author} {\bibfnamefont {C.}~\bibnamefont
  {Lled\'o}}, \bibinfo {author} {\bibfnamefont {T.~K.}\ \bibnamefont
  {Mavrogordatos}},\ and\ \bibinfo {author} {\bibfnamefont {M.~H.}\
  \bibnamefont {Szyma\ifmmode~\acute{n}\else \'{n}\fi{}ska}},\ }\href
  {https://doi.org/10.1103/PhysRevB.100.054303} {\bibfield  {journal} {\bibinfo
   {journal} {Phys. Rev. B}\ }\textbf {\bibinfo {volume} {100}},\ \bibinfo
  {pages} {054303} (\bibinfo {year} {2019})}\BibitemShut {NoStop}%
\bibitem [{\citenamefont {Tucker}\ \emph {et~al.}(2018)\citenamefont {Tucker},
  \citenamefont {Zhu}, \citenamefont {Lewis-Swan}, \citenamefont {Marino},
  \citenamefont {Jimenez}, \citenamefont {Restrepo},\ and\ \citenamefont
  {Rey}}]{Jamir1}%
  \BibitemOpen
  \bibfield  {author} {\bibinfo {author} {\bibfnamefont {K.}~\bibnamefont
  {Tucker}}, \bibinfo {author} {\bibfnamefont {B.}~\bibnamefont {Zhu}},
  \bibinfo {author} {\bibfnamefont {R.~J.}\ \bibnamefont {Lewis-Swan}},
  \bibinfo {author} {\bibfnamefont {J.}~\bibnamefont {Marino}}, \bibinfo
  {author} {\bibfnamefont {F.}~\bibnamefont {Jimenez}}, \bibinfo {author}
  {\bibfnamefont {J.~G.}\ \bibnamefont {Restrepo}},\ and\ \bibinfo {author}
  {\bibfnamefont {A.~M.}\ \bibnamefont {Rey}},\ }\href@noop {} {\bibfield
  {journal} {\bibinfo  {journal} {New Journal of Physics}\ }\textbf {\bibinfo
  {volume} {20}},\ \bibinfo {pages} {123003} (\bibinfo {year}
  {2018})}\BibitemShut {NoStop}%
\bibitem [{\citenamefont {Zhu}\ \emph {et~al.}(2019)\citenamefont {Zhu},
  \citenamefont {Marino}, \citenamefont {Yao}, \citenamefont {Lukin},\ and\
  \citenamefont {Demler}}]{Jamir2}%
  \BibitemOpen
  \bibfield  {author} {\bibinfo {author} {\bibfnamefont {B.}~\bibnamefont
  {Zhu}}, \bibinfo {author} {\bibfnamefont {J.}~\bibnamefont {Marino}},
  \bibinfo {author} {\bibfnamefont {N.~Y.}\ \bibnamefont {Yao}}, \bibinfo
  {author} {\bibfnamefont {M.~D.}\ \bibnamefont {Lukin}},\ and\ \bibinfo
  {author} {\bibfnamefont {E.~A.}\ \bibnamefont {Demler}},\ }\href@noop {}
  {\bibfield  {journal} {\bibinfo  {journal} {New Journal of Physics}\ }\textbf
  {\bibinfo {volume} {21}},\ \bibinfo {pages} {073028} (\bibinfo {year}
  {2019})}\BibitemShut {NoStop}%
\bibitem [{\citenamefont {Chinzei}\ and\ \citenamefont
  {Ikeda}(2020)}]{Chinzei}%
  \BibitemOpen
  \bibfield  {author} {\bibinfo {author} {\bibfnamefont {K.}~\bibnamefont
  {Chinzei}}\ and\ \bibinfo {author} {\bibfnamefont {T.~N.}\ \bibnamefont
  {Ikeda}},\ }\href {https://doi.org/10.1103/PhysRevLett.125.060601} {\bibfield
   {journal} {\bibinfo  {journal} {Phys. Rev. Lett.}\ }\textbf {\bibinfo
  {volume} {125}},\ \bibinfo {pages} {060601} (\bibinfo {year}
  {2020})}\BibitemShut {NoStop}%
\bibitem [{\citenamefont {Cosme}\ \emph {et~al.}(2019)\citenamefont {Cosme},
  \citenamefont {Skulte},\ and\ \citenamefont {Mathey}}]{Cosme1}%
  \BibitemOpen
  \bibfield  {author} {\bibinfo {author} {\bibfnamefont {J.~G.}\ \bibnamefont
  {Cosme}}, \bibinfo {author} {\bibfnamefont {J.}~\bibnamefont {Skulte}},\ and\
  \bibinfo {author} {\bibfnamefont {L.}~\bibnamefont {Mathey}},\ }\href
  {https://doi.org/10.1103/PhysRevA.100.053615} {\bibfield  {journal} {\bibinfo
   {journal} {Phys. Rev. A}\ }\textbf {\bibinfo {volume} {100}},\ \bibinfo
  {pages} {053615} (\bibinfo {year} {2019})}\BibitemShut {NoStop}%
\bibitem [{\citenamefont {Dogra}\ \emph {et~al.}(2019)\citenamefont {Dogra},
  \citenamefont {Landini}, \citenamefont {Kroeger}, \citenamefont {Hruby},
  \citenamefont {Donner},\ and\ \citenamefont {Esslinger}}]{Dogra1496}%
  \BibitemOpen
  \bibfield  {author} {\bibinfo {author} {\bibfnamefont {N.}~\bibnamefont
  {Dogra}}, \bibinfo {author} {\bibfnamefont {M.}~\bibnamefont {Landini}},
  \bibinfo {author} {\bibfnamefont {K.}~\bibnamefont {Kroeger}}, \bibinfo
  {author} {\bibfnamefont {L.}~\bibnamefont {Hruby}}, \bibinfo {author}
  {\bibfnamefont {T.}~\bibnamefont {Donner}},\ and\ \bibinfo {author}
  {\bibfnamefont {T.}~\bibnamefont {Esslinger}},\ }\href
  {https://doi.org/10.1126/science.aaw4465} {\bibfield  {journal} {\bibinfo
  {journal} {Science}\ }\textbf {\bibinfo {volume} {366}},\ \bibinfo {pages}
  {1496} (\bibinfo {year} {2019})},\ \Eprint
  {https://arxiv.org/abs/https://science.sciencemag.org/content/366/6472/1496.full.pdf}
  {https://science.sciencemag.org/content/366/6472/1496.full.pdf} \BibitemShut
  {NoStop}%
\bibitem [{\citenamefont {Ke\ss{}ler}\ \emph {et~al.}(2021)\citenamefont
  {Ke\ss{}ler}, \citenamefont {Kongkhambut}, \citenamefont {Georges},
  \citenamefont {Mathey}, \citenamefont {Cosme},\ and\ \citenamefont
  {Hemmerich}}]{dissipativeTCobs}%
  \BibitemOpen
  \bibfield  {author} {\bibinfo {author} {\bibfnamefont {H.}~\bibnamefont
  {Ke\ss{}ler}}, \bibinfo {author} {\bibfnamefont {P.}~\bibnamefont
  {Kongkhambut}}, \bibinfo {author} {\bibfnamefont {C.}~\bibnamefont
  {Georges}}, \bibinfo {author} {\bibfnamefont {L.}~\bibnamefont {Mathey}},
  \bibinfo {author} {\bibfnamefont {J.~G.}\ \bibnamefont {Cosme}},\ and\
  \bibinfo {author} {\bibfnamefont {A.}~\bibnamefont {Hemmerich}},\ }\href
  {https://doi.org/10.1103/PhysRevLett.127.043602} {\bibfield  {journal}
  {\bibinfo  {journal} {Phys. Rev. Lett.}\ }\textbf {\bibinfo {volume} {127}},\
  \bibinfo {pages} {043602} (\bibinfo {year} {2021})}\BibitemShut {NoStop}%
\bibitem [{\citenamefont {Kongkhambut}\ \emph {et~al.}(2022)\citenamefont
  {Kongkhambut}, \citenamefont {Skulte}, \citenamefont {Mathey}, \citenamefont
  {Cosme}, \citenamefont {Hemmerich},\ and\ \citenamefont
  {Keßler}}]{DissipativeContObs}%
  \BibitemOpen
  \bibfield  {author} {\bibinfo {author} {\bibfnamefont {P.}~\bibnamefont
  {Kongkhambut}}, \bibinfo {author} {\bibfnamefont {J.}~\bibnamefont {Skulte}},
  \bibinfo {author} {\bibfnamefont {L.}~\bibnamefont {Mathey}}, \bibinfo
  {author} {\bibfnamefont {J.~G.}\ \bibnamefont {Cosme}}, \bibinfo {author}
  {\bibfnamefont {A.}~\bibnamefont {Hemmerich}},\ and\ \bibinfo {author}
  {\bibfnamefont {H.}~\bibnamefont {Keßler}},\ }\href@noop {} {\bibinfo
  {title} {Observation of a continuous time crystal}} (\bibinfo {year}
  {2022}),\ \Eprint {https://arxiv.org/abs/2202.06980} {arXiv:2202.06980
  [cond-mat.quant-gas]} \BibitemShut {NoStop}%
\bibitem [{\citenamefont {Dreon}\ \emph {et~al.}(2021)\citenamefont {Dreon},
  \citenamefont {Baumg{\"a}rtner}, \citenamefont {Li}, \citenamefont
  {Hertlein}, \citenamefont {Esslinger},\ and\ \citenamefont
  {Donner}}]{dreon2021self}%
  \BibitemOpen
  \bibfield  {author} {\bibinfo {author} {\bibfnamefont {D.}~\bibnamefont
  {Dreon}}, \bibinfo {author} {\bibfnamefont {A.}~\bibnamefont
  {Baumg{\"a}rtner}}, \bibinfo {author} {\bibfnamefont {X.}~\bibnamefont {Li}},
  \bibinfo {author} {\bibfnamefont {S.}~\bibnamefont {Hertlein}}, \bibinfo
  {author} {\bibfnamefont {T.}~\bibnamefont {Esslinger}},\ and\ \bibinfo
  {author} {\bibfnamefont {T.}~\bibnamefont {Donner}},\ }\href@noop {}
  {\bibfield  {journal} {\bibinfo  {journal} {arXiv preprint arXiv:2112.11502}\
  } (\bibinfo {year} {2021})}\BibitemShut {NoStop}%
\bibitem [{\citenamefont {Klinder}\ \emph {et~al.}(2015)\citenamefont
  {Klinder}, \citenamefont {Keßler}, \citenamefont {Wolke}, \citenamefont
  {Mathey},\ and\ \citenamefont {Hemmerich}}]{HemmerichPNAS}%
  \BibitemOpen
  \bibfield  {author} {\bibinfo {author} {\bibfnamefont {J.}~\bibnamefont
  {Klinder}}, \bibinfo {author} {\bibfnamefont {H.}~\bibnamefont {Keßler}},
  \bibinfo {author} {\bibfnamefont {M.}~\bibnamefont {Wolke}}, \bibinfo
  {author} {\bibfnamefont {L.}~\bibnamefont {Mathey}},\ and\ \bibinfo {author}
  {\bibfnamefont {A.}~\bibnamefont {Hemmerich}},\ }\href
  {https://doi.org/10.1073/pnas.1417132112} {\bibfield  {journal} {\bibinfo
  {journal} {Proceedings of the National Academy of Sciences}\ }\textbf
  {\bibinfo {volume} {112}},\ \bibinfo {pages} {3290} (\bibinfo {year}
  {2015})},\ \Eprint
  {https://arxiv.org/abs/https://www.pnas.org/doi/pdf/10.1073/pnas.1417132112}
  {https://www.pnas.org/doi/pdf/10.1073/pnas.1417132112} \BibitemShut {NoStop}%
\bibitem [{\citenamefont {Kollár}\ \emph {et~al.}(2017)\citenamefont
  {Kollár}, \citenamefont {Papageorge}, \citenamefont {Vaidya}, \citenamefont
  {Guo}, \citenamefont {Keeling},\ and\ \citenamefont {Lev}}]{LevNatComms}%
  \BibitemOpen
  \bibfield  {author} {\bibinfo {author} {\bibfnamefont {A.~J.}\ \bibnamefont
  {Kollár}}, \bibinfo {author} {\bibfnamefont {A.~T.}\ \bibnamefont
  {Papageorge}}, \bibinfo {author} {\bibfnamefont {V.~D.}\ \bibnamefont
  {Vaidya}}, \bibinfo {author} {\bibfnamefont {Y.}~\bibnamefont {Guo}},
  \bibinfo {author} {\bibfnamefont {J.}~\bibnamefont {Keeling}},\ and\ \bibinfo
  {author} {\bibfnamefont {B.~L.}\ \bibnamefont {Lev}},\ }\bibfield  {journal}
  {\bibinfo  {journal} {Nature Communications}\ }\textbf {\bibinfo {volume}
  {8}},\ \href {https://doi.org/10.1038/ncomms14386} {10.1038/ncomms14386}
  (\bibinfo {year} {2017})\BibitemShut {NoStop}%
\bibitem [{\citenamefont {Baumann}\ \emph
  {et~al.}(2010{\natexlab{b}})\citenamefont {Baumann}, \citenamefont {Guerlin},
  \citenamefont {Brennecke},\ and\ \citenamefont
  {Esslinger}}]{baumann2010dicke}%
  \BibitemOpen
  \bibfield  {author} {\bibinfo {author} {\bibfnamefont {K.}~\bibnamefont
  {Baumann}}, \bibinfo {author} {\bibfnamefont {C.}~\bibnamefont {Guerlin}},
  \bibinfo {author} {\bibfnamefont {F.}~\bibnamefont {Brennecke}},\ and\
  \bibinfo {author} {\bibfnamefont {T.}~\bibnamefont {Esslinger}},\ }\href@noop
  {} {\bibfield  {journal} {\bibinfo  {journal} {nature}\ }\textbf {\bibinfo
  {volume} {464}},\ \bibinfo {pages} {1301} (\bibinfo {year}
  {2010}{\natexlab{b}})}\BibitemShut {NoStop}%
\bibitem [{\citenamefont {{Zhang}}\ \emph {et~al.}(2021)\citenamefont
  {{Zhang}}, \citenamefont {{Chen}}, \citenamefont {{Wu}}, \citenamefont
  {{Wang}}, \citenamefont {{Fan}}, \citenamefont {{Deng}},\ and\ \citenamefont
  {{Wu}}}]{ZhangScience}%
  \BibitemOpen
  \bibfield  {author} {\bibinfo {author} {\bibfnamefont {X.}~\bibnamefont
  {{Zhang}}}, \bibinfo {author} {\bibfnamefont {Y.}~\bibnamefont {{Chen}}},
  \bibinfo {author} {\bibfnamefont {Z.}~\bibnamefont {{Wu}}}, \bibinfo {author}
  {\bibfnamefont {J.}~\bibnamefont {{Wang}}}, \bibinfo {author} {\bibfnamefont
  {J.}~\bibnamefont {{Fan}}}, \bibinfo {author} {\bibfnamefont
  {S.}~\bibnamefont {{Deng}}},\ and\ \bibinfo {author} {\bibfnamefont
  {H.}~\bibnamefont {{Wu}}},\ }\href {https://doi.org/10.1126/science.abd4385}
  {\bibfield  {journal} {\bibinfo  {journal} {Science}\ }\textbf {\bibinfo
  {volume} {373}},\ \bibinfo {pages} {1359} (\bibinfo {year}
  {2021})}\BibitemShut {NoStop}%
\bibitem [{\citenamefont {Mivehvar}\ \emph
  {et~al.}(2021{\natexlab{b}})\citenamefont {Mivehvar}, \citenamefont {Piazza},
  \citenamefont {Donner},\ and\ \citenamefont {Ritsch}}]{mivehvar2021cavity}%
  \BibitemOpen
  \bibfield  {author} {\bibinfo {author} {\bibfnamefont {F.}~\bibnamefont
  {Mivehvar}}, \bibinfo {author} {\bibfnamefont {F.}~\bibnamefont {Piazza}},
  \bibinfo {author} {\bibfnamefont {T.}~\bibnamefont {Donner}},\ and\ \bibinfo
  {author} {\bibfnamefont {H.}~\bibnamefont {Ritsch}},\ }\href@noop {}
  {\bibfield  {journal} {\bibinfo  {journal} {arXiv preprint arXiv:2102.04473}\
  } (\bibinfo {year} {2021}{\natexlab{b}})}\BibitemShut {NoStop}%
\bibitem [{\citenamefont {Skulte}\ \emph {et~al.}(2021)\citenamefont {Skulte},
  \citenamefont {Kongkhambut}, \citenamefont {Keßler}, \citenamefont
  {Hemmerich}, \citenamefont {Mathey},\ and\ \citenamefont {Cosme}}]{3Dicke1}%
  \BibitemOpen
  \bibfield  {author} {\bibinfo {author} {\bibfnamefont {J.}~\bibnamefont
  {Skulte}}, \bibinfo {author} {\bibfnamefont {P.}~\bibnamefont {Kongkhambut}},
  \bibinfo {author} {\bibfnamefont {H.}~\bibnamefont {Keßler}}, \bibinfo
  {author} {\bibfnamefont {A.}~\bibnamefont {Hemmerich}}, \bibinfo {author}
  {\bibfnamefont {L.}~\bibnamefont {Mathey}},\ and\ \bibinfo {author}
  {\bibfnamefont {J.~G.}\ \bibnamefont {Cosme}},\ }\bibfield  {journal}
  {\bibinfo  {journal} {Physical Review A}\ }\textbf {\bibinfo {volume}
  {104}},\ \href {https://doi.org/10.1103/physreva.104.063705}
  {10.1103/physreva.104.063705} (\bibinfo {year} {2021})\BibitemShut {NoStop}%
\bibitem [{\citenamefont {Cola}\ \emph {et~al.}(2009)\citenamefont {Cola},
  \citenamefont {Bigerni},\ and\ \citenamefont {Piovella}}]{3Dicke2}%
  \BibitemOpen
  \bibfield  {author} {\bibinfo {author} {\bibfnamefont {M.~M.}\ \bibnamefont
  {Cola}}, \bibinfo {author} {\bibfnamefont {D.}~\bibnamefont {Bigerni}},\ and\
  \bibinfo {author} {\bibfnamefont {N.}~\bibnamefont {Piovella}},\ }\href
  {https://doi.org/10.1103/PhysRevA.79.053622} {\bibfield  {journal} {\bibinfo
  {journal} {Phys. Rev. A}\ }\textbf {\bibinfo {volume} {79}},\ \bibinfo
  {pages} {053622} (\bibinfo {year} {2009})}\BibitemShut {NoStop}%
\bibitem [{\citenamefont {Fan}\ \emph {et~al.}(2020)\citenamefont {Fan},
  \citenamefont {Chen},\ and\ \citenamefont {Jia}}]{3dicke3}%
  \BibitemOpen
  \bibfield  {author} {\bibinfo {author} {\bibfnamefont {J.}~\bibnamefont
  {Fan}}, \bibinfo {author} {\bibfnamefont {G.}~\bibnamefont {Chen}},\ and\
  \bibinfo {author} {\bibfnamefont {S.}~\bibnamefont {Jia}},\ }\bibfield
  {journal} {\bibinfo  {journal} {Physical Review A}\ }\textbf {\bibinfo
  {volume} {101}},\ \href {https://doi.org/10.1103/physreva.101.063627}
  {10.1103/physreva.101.063627} (\bibinfo {year} {2020})\BibitemShut {NoStop}%
\bibitem [{\citenamefont {Kirton}\ \emph {et~al.}(2018)\citenamefont {Kirton},
  \citenamefont {Roses}, \citenamefont {Keeling},\ and\ \citenamefont
  {Dalla~Torre}}]{Dickereview}%
  \BibitemOpen
  \bibfield  {author} {\bibinfo {author} {\bibfnamefont {P.}~\bibnamefont
  {Kirton}}, \bibinfo {author} {\bibfnamefont {M.~M.}\ \bibnamefont {Roses}},
  \bibinfo {author} {\bibfnamefont {J.}~\bibnamefont {Keeling}},\ and\ \bibinfo
  {author} {\bibfnamefont {E.~G.}\ \bibnamefont {Dalla~Torre}},\ }\href
  {https://doi.org/10.1002/qute.201800043} {\bibfield  {journal} {\bibinfo
  {journal} {Advanced Quantum Technologies}\ }\textbf {\bibinfo {volume} {2}},\
  \bibinfo {pages} {1800043} (\bibinfo {year} {2018})}\BibitemShut {NoStop}%
\bibitem [{\citenamefont {Zheng}\ and\ \citenamefont
  {Cooper}(2018)}]{zheng2018anomalous}%
  \BibitemOpen
  \bibfield  {author} {\bibinfo {author} {\bibfnamefont {W.}~\bibnamefont
  {Zheng}}\ and\ \bibinfo {author} {\bibfnamefont {N.~R.}\ \bibnamefont
  {Cooper}},\ }\href@noop {} {\bibfield  {journal} {\bibinfo  {journal}
  {Physical Review A}\ }\textbf {\bibinfo {volume} {97}},\ \bibinfo {pages}
  {021601} (\bibinfo {year} {2018})}\BibitemShut {NoStop}%
\bibitem [{\citenamefont {Meinert}\ \emph {et~al.}(2017)\citenamefont
  {Meinert}, \citenamefont {Knap}, \citenamefont {Kirilov}, \citenamefont
  {Jag-Lauber}, \citenamefont {Zvonarev}, \citenamefont {Demler},\ and\
  \citenamefont {N{\"a}gerl}}]{BlochScience}%
  \BibitemOpen
  \bibfield  {author} {\bibinfo {author} {\bibfnamefont {F.}~\bibnamefont
  {Meinert}}, \bibinfo {author} {\bibfnamefont {M.}~\bibnamefont {Knap}},
  \bibinfo {author} {\bibfnamefont {E.}~\bibnamefont {Kirilov}}, \bibinfo
  {author} {\bibfnamefont {K.}~\bibnamefont {Jag-Lauber}}, \bibinfo {author}
  {\bibfnamefont {M.~B.}\ \bibnamefont {Zvonarev}}, \bibinfo {author}
  {\bibfnamefont {E.}~\bibnamefont {Demler}},\ and\ \bibinfo {author}
  {\bibfnamefont {H.-C.}\ \bibnamefont {N{\"a}gerl}},\ }\href
  {https://doi.org/10.1126/science.aah6616} {\bibfield  {journal} {\bibinfo
  {journal} {Science}\ }\textbf {\bibinfo {volume} {356}},\ \bibinfo {pages}
  {945} (\bibinfo {year} {2017})},\ \Eprint
  {https://arxiv.org/abs/https://science.sciencemag.org/content/356/6341/945.full.pdf}
  {https://science.sciencemag.org/content/356/6341/945.full.pdf} \BibitemShut
  {NoStop}%
\bibitem [{SM(2022)}]{SM}%
  \BibitemOpen
  \href@noop {} {}\bibinfo {howpublished} {Supplemental Material} (\bibinfo
  {year} {2022})\BibitemShut {NoStop}%
\bibitem [{\citenamefont {Tindall}\ \emph {et~al.}(2020)\citenamefont
  {Tindall}, \citenamefont {Sánchez~Muñoz}, \citenamefont {Buča},\ and\
  \citenamefont {Jaksch}}]{quantumsynch}%
  \BibitemOpen
  \bibfield  {author} {\bibinfo {author} {\bibfnamefont {J.}~\bibnamefont
  {Tindall}}, \bibinfo {author} {\bibfnamefont {C.}~\bibnamefont
  {Sánchez~Muñoz}}, \bibinfo {author} {\bibfnamefont {B.}~\bibnamefont
  {Buča}},\ and\ \bibinfo {author} {\bibfnamefont {D.}~\bibnamefont
  {Jaksch}},\ }\href {https://doi.org/10.1088/1367-2630/ab60f5} {\bibfield
  {journal} {\bibinfo  {journal} {New Journal of Physics}\ }\textbf {\bibinfo
  {volume} {22}},\ \bibinfo {pages} {013026} (\bibinfo {year}
  {2020})}\BibitemShut {NoStop}%
\bibitem [{\citenamefont {Buca}\ \emph {et~al.}(2021)\citenamefont {Buca},
  \citenamefont {Booker},\ and\ \citenamefont {Jaksch}}]{buca2021algebraic}%
  \BibitemOpen
  \bibfield  {author} {\bibinfo {author} {\bibfnamefont {B.}~\bibnamefont
  {Buca}}, \bibinfo {author} {\bibfnamefont {C.}~\bibnamefont {Booker}},\ and\
  \bibinfo {author} {\bibfnamefont {D.}~\bibnamefont {Jaksch}},\ }\href@noop {}
  {\bibinfo {title} {Algebraic theory of quantum synchronization and limit
  cycles under dissipation}} (\bibinfo {year} {2021}),\ \Eprint
  {https://arxiv.org/abs/2103.01808} {arXiv:2103.01808 [quant-ph]} \BibitemShut
  {NoStop}%
\bibitem [{\citenamefont {Medenjak}\ \emph {et~al.}(2020)\citenamefont
  {Medenjak}, \citenamefont {Bu\ifmmode~\check{c}\else \v{c}\fi{}a},\ and\
  \citenamefont {Jaksch}}]{Marko1}%
  \BibitemOpen
  \bibfield  {author} {\bibinfo {author} {\bibfnamefont {M.}~\bibnamefont
  {Medenjak}}, \bibinfo {author} {\bibfnamefont {B.}~\bibnamefont
  {Bu\ifmmode~\check{c}\else \v{c}\fi{}a}},\ and\ \bibinfo {author}
  {\bibfnamefont {D.}~\bibnamefont {Jaksch}},\ }\href
  {https://doi.org/10.1103/PhysRevB.102.041117} {\bibfield  {journal} {\bibinfo
   {journal} {Phys. Rev. B}\ }\textbf {\bibinfo {volume} {102}},\ \bibinfo
  {pages} {041117} (\bibinfo {year} {2020})}\BibitemShut {NoStop}%
\bibitem [{\citenamefont {Battesti}\ \emph {et~al.}(2004)\citenamefont
  {Battesti}, \citenamefont {Clad{\'e}}, \citenamefont {Guellati-Kh{\'e}lifa},
  \citenamefont {Schwob}, \citenamefont {Gr{\'e}maud}, \citenamefont {Nez},
  \citenamefont {Julien},\ and\ \citenamefont {Biraben}}]{battesti2004bloch}%
  \BibitemOpen
  \bibfield  {author} {\bibinfo {author} {\bibfnamefont {R.}~\bibnamefont
  {Battesti}}, \bibinfo {author} {\bibfnamefont {P.}~\bibnamefont {Clad{\'e}}},
  \bibinfo {author} {\bibfnamefont {S.}~\bibnamefont {Guellati-Kh{\'e}lifa}},
  \bibinfo {author} {\bibfnamefont {C.}~\bibnamefont {Schwob}}, \bibinfo
  {author} {\bibfnamefont {B.}~\bibnamefont {Gr{\'e}maud}}, \bibinfo {author}
  {\bibfnamefont {F.}~\bibnamefont {Nez}}, \bibinfo {author} {\bibfnamefont
  {L.}~\bibnamefont {Julien}},\ and\ \bibinfo {author} {\bibfnamefont
  {F.}~\bibnamefont {Biraben}},\ }\href@noop {} {\bibfield  {journal} {\bibinfo
   {journal} {Physical review letters}\ }\textbf {\bibinfo {volume} {92}},\
  \bibinfo {pages} {253001} (\bibinfo {year} {2004})}\BibitemShut {NoStop}%
\bibitem [{\citenamefont {Dahan}\ \emph {et~al.}(1996)\citenamefont {Dahan},
  \citenamefont {Peik}, \citenamefont {Reichel}, \citenamefont {Castin},\ and\
  \citenamefont {Salomon}}]{dahan1996bloch}%
  \BibitemOpen
  \bibfield  {author} {\bibinfo {author} {\bibfnamefont {M.~B.}\ \bibnamefont
  {Dahan}}, \bibinfo {author} {\bibfnamefont {E.}~\bibnamefont {Peik}},
  \bibinfo {author} {\bibfnamefont {J.}~\bibnamefont {Reichel}}, \bibinfo
  {author} {\bibfnamefont {Y.}~\bibnamefont {Castin}},\ and\ \bibinfo {author}
  {\bibfnamefont {C.}~\bibnamefont {Salomon}},\ }\href@noop {} {\bibfield
  {journal} {\bibinfo  {journal} {Physical Review Letters}\ }\textbf {\bibinfo
  {volume} {76}},\ \bibinfo {pages} {4508} (\bibinfo {year}
  {1996})}\BibitemShut {NoStop}%
\bibitem [{\citenamefont {Ljubotina}\ \emph {et~al.}(2017)\citenamefont
  {Ljubotina}, \citenamefont {{\v{Z}}nidari{\v{c}}},\ and\ \citenamefont
  {Prosen}}]{ljubotina2017spin}%
  \BibitemOpen
  \bibfield  {author} {\bibinfo {author} {\bibfnamefont {M.}~\bibnamefont
  {Ljubotina}}, \bibinfo {author} {\bibfnamefont {M.}~\bibnamefont
  {{\v{Z}}nidari{\v{c}}}},\ and\ \bibinfo {author} {\bibfnamefont
  {T.}~\bibnamefont {Prosen}},\ }\href@noop {} {\bibfield  {journal} {\bibinfo
  {journal} {Nature communications}\ }\textbf {\bibinfo {volume} {8}},\
  \bibinfo {pages} {1} (\bibinfo {year} {2017})}\BibitemShut {NoStop}%
\bibitem [{\citenamefont {Ilievski}\ \emph {et~al.}(2018)\citenamefont
  {Ilievski}, \citenamefont {De~Nardis}, \citenamefont {Medenjak},\ and\
  \citenamefont {Prosen}}]{PhysRevLett.121.230602}%
  \BibitemOpen
  \bibfield  {author} {\bibinfo {author} {\bibfnamefont {E.}~\bibnamefont
  {Ilievski}}, \bibinfo {author} {\bibfnamefont {J.}~\bibnamefont {De~Nardis}},
  \bibinfo {author} {\bibfnamefont {M.}~\bibnamefont {Medenjak}},\ and\
  \bibinfo {author} {\bibfnamefont {T.~c.~v.}\ \bibnamefont {Prosen}},\ }\href
  {https://doi.org/10.1103/PhysRevLett.121.230602} {\bibfield  {journal}
  {\bibinfo  {journal} {Phys. Rev. Lett.}\ }\textbf {\bibinfo {volume} {121}},\
  \bibinfo {pages} {230602} (\bibinfo {year} {2018})}\BibitemShut {NoStop}%
\bibitem [{\citenamefont {Cronin}\ \emph {et~al.}(2009)\citenamefont {Cronin},
  \citenamefont {Schmiedmayer},\ and\ \citenamefont
  {Pritchard}}]{cronin2009optics}%
  \BibitemOpen
  \bibfield  {author} {\bibinfo {author} {\bibfnamefont {A.~D.}\ \bibnamefont
  {Cronin}}, \bibinfo {author} {\bibfnamefont {J.}~\bibnamefont
  {Schmiedmayer}},\ and\ \bibinfo {author} {\bibfnamefont {D.~E.}\ \bibnamefont
  {Pritchard}},\ }\href@noop {} {\bibfield  {journal} {\bibinfo  {journal}
  {Reviews of Modern Physics}\ }\textbf {\bibinfo {volume} {81}},\ \bibinfo
  {pages} {1051} (\bibinfo {year} {2009})}\BibitemShut {NoStop}%
\bibitem [{\citenamefont {Chiow}\ \emph {et~al.}(2011)\citenamefont {Chiow},
  \citenamefont {Kovachy}, \citenamefont {Chien},\ and\ \citenamefont
  {Kasevich}}]{chiow2011102}%
  \BibitemOpen
  \bibfield  {author} {\bibinfo {author} {\bibfnamefont {S.-w.}\ \bibnamefont
  {Chiow}}, \bibinfo {author} {\bibfnamefont {T.}~\bibnamefont {Kovachy}},
  \bibinfo {author} {\bibfnamefont {H.-C.}\ \bibnamefont {Chien}},\ and\
  \bibinfo {author} {\bibfnamefont {M.~A.}\ \bibnamefont {Kasevich}},\
  }\href@noop {} {\bibfield  {journal} {\bibinfo  {journal} {Physical review
  letters}\ }\textbf {\bibinfo {volume} {107}},\ \bibinfo {pages} {130403}
  (\bibinfo {year} {2011})}\BibitemShut {NoStop}%
\bibitem [{\citenamefont {Hohenester}\ \emph {et~al.}(2007)\citenamefont
  {Hohenester}, \citenamefont {Rekdal}, \citenamefont {Borzi},\ and\
  \citenamefont {Schmiedmayer}}]{hohenester2007optimal}%
  \BibitemOpen
  \bibfield  {author} {\bibinfo {author} {\bibfnamefont {U.}~\bibnamefont
  {Hohenester}}, \bibinfo {author} {\bibfnamefont {P.~K.}\ \bibnamefont
  {Rekdal}}, \bibinfo {author} {\bibfnamefont {A.}~\bibnamefont {Borzi}},\ and\
  \bibinfo {author} {\bibfnamefont {J.}~\bibnamefont {Schmiedmayer}},\
  }\href@noop {} {\bibfield  {journal} {\bibinfo  {journal} {Physical Review
  A}\ }\textbf {\bibinfo {volume} {75}},\ \bibinfo {pages} {023602} (\bibinfo
  {year} {2007})}\BibitemShut {NoStop}%
\bibitem [{\citenamefont {Azouit}\ \emph {et~al.}(2016)\citenamefont {Azouit},
  \citenamefont {Sarlette},\ and\ \citenamefont
  {Rouchon}}]{azouit2016adiabatic}%
  \BibitemOpen
  \bibfield  {author} {\bibinfo {author} {\bibfnamefont {R.}~\bibnamefont
  {Azouit}}, \bibinfo {author} {\bibfnamefont {A.}~\bibnamefont {Sarlette}},\
  and\ \bibinfo {author} {\bibfnamefont {P.}~\bibnamefont {Rouchon}},\ }in\
  \href@noop {} {\emph {\bibinfo {booktitle} {2016 IEEE 55th Conference on
  Decision and Control (CDC)}}}\ (\bibinfo {organization} {IEEE},\ \bibinfo
  {year} {2016})\ pp.\ \bibinfo {pages} {4559--4565}\BibitemShut {NoStop}%
\bibitem [{\citenamefont {Lacroix}\ \emph {et~al.}(2011)\citenamefont
  {Lacroix}, \citenamefont {Mendels},\ and\ \citenamefont
  {Mila}}]{lacroix2011introduction}%
  \BibitemOpen
  \bibfield  {author} {\bibinfo {author} {\bibfnamefont {C.}~\bibnamefont
  {Lacroix}}, \bibinfo {author} {\bibfnamefont {P.}~\bibnamefont {Mendels}},\
  and\ \bibinfo {author} {\bibfnamefont {F.}~\bibnamefont {Mila}},\ }\href@noop
  {} {\emph {\bibinfo {title} {Introduction to frustrated magnetism: materials,
  experiments, theory}}},\ Vol.\ \bibinfo {volume} {164}\ (\bibinfo
  {publisher} {Springer Science \& Business Media},\ \bibinfo {year}
  {2011})\BibitemShut {NoStop}%
\bibitem [{\citenamefont {Wagner}(1975)}]{wagner1975nonlinear}%
  \BibitemOpen
  \bibfield  {author} {\bibinfo {author} {\bibfnamefont {M.}~\bibnamefont
  {Wagner}},\ }\href@noop {} {\bibfield  {journal} {\bibinfo  {journal}
  {Physics Letters A}\ }\textbf {\bibinfo {volume} {53}},\ \bibinfo {pages} {1}
  (\bibinfo {year} {1975})}\BibitemShut {NoStop}%
\bibitem [{\citenamefont {Prosen}(2008)}]{Prosen_2008}%
  \BibitemOpen
  \bibfield  {author} {\bibinfo {author} {\bibfnamefont {T.}~\bibnamefont
  {Prosen}},\ }\href {https://doi.org/10.1088/1367-2630/10/4/043026} {\bibfield
   {journal} {\bibinfo  {journal} {New Journal of Physics}\ }\textbf {\bibinfo
  {volume} {10}},\ \bibinfo {pages} {043026} (\bibinfo {year}
  {2008})}\BibitemShut {NoStop}%
\bibitem [{\citenamefont {Prosen}\ and\ \citenamefont
  {Seligman}(2010)}]{prosen2010quantization}%
  \BibitemOpen
  \bibfield  {author} {\bibinfo {author} {\bibfnamefont {T.}~\bibnamefont
  {Prosen}}\ and\ \bibinfo {author} {\bibfnamefont {T.~H.}\ \bibnamefont
  {Seligman}},\ }\href@noop {} {\bibinfo {title} {Quantization over boson
  operator spaces}} (\bibinfo {year} {2010}),\ \Eprint
  {https://arxiv.org/abs/1007.2921} {arXiv:1007.2921 [quant-ph]} \BibitemShut
  {NoStop}%
\bibitem [{\citenamefont {Fr\"oml}\ \emph {et~al.}(2020)\citenamefont
  {Fr\"oml}, \citenamefont {Muckel}, \citenamefont {Kollath}, \citenamefont
  {Chiocchetta},\ and\ \citenamefont {Diehl}}]{Zeno1}%
  \BibitemOpen
  \bibfield  {author} {\bibinfo {author} {\bibfnamefont {H.}~\bibnamefont
  {Fr\"oml}}, \bibinfo {author} {\bibfnamefont {C.}~\bibnamefont {Muckel}},
  \bibinfo {author} {\bibfnamefont {C.}~\bibnamefont {Kollath}}, \bibinfo
  {author} {\bibfnamefont {A.}~\bibnamefont {Chiocchetta}},\ and\ \bibinfo
  {author} {\bibfnamefont {S.}~\bibnamefont {Diehl}},\ }\href
  {https://doi.org/10.1103/PhysRevB.101.144301} {\bibfield  {journal} {\bibinfo
   {journal} {Phys. Rev. B}\ }\textbf {\bibinfo {volume} {101}},\ \bibinfo
  {pages} {144301} (\bibinfo {year} {2020})}\BibitemShut {NoStop}%
\bibitem [{\citenamefont {Bezvershenko}\ \emph {et~al.}(2020)\citenamefont
  {Bezvershenko}, \citenamefont {Halati}, \citenamefont {Sheikhan},
  \citenamefont {Kollath},\ and\ \citenamefont
  {Rosch}}]{bezvershenko2020dicke}%
  \BibitemOpen
  \bibfield  {author} {\bibinfo {author} {\bibfnamefont {A.~V.}\ \bibnamefont
  {Bezvershenko}}, \bibinfo {author} {\bibfnamefont {C.-M.}\ \bibnamefont
  {Halati}}, \bibinfo {author} {\bibfnamefont {A.}~\bibnamefont {Sheikhan}},
  \bibinfo {author} {\bibfnamefont {C.}~\bibnamefont {Kollath}},\ and\ \bibinfo
  {author} {\bibfnamefont {A.}~\bibnamefont {Rosch}},\ }\href@noop {}
  {\bibfield  {journal} {\bibinfo  {journal} {arXiv preprint arXiv:2012.11823}\
  } (\bibinfo {year} {2020})}\BibitemShut {NoStop}%
\end{thebibliography}%
\widetext
\clearpage
\begin{center}
\textbf{\large Supplementary Material: Tunable Phase Transitions between Spatial and Temporal Order through Dissipation}\\
\end{center}
\setcounter{equation}{0}
\setcounter{figure}{0}
\setcounter{table}{0}
\setcounter{page}{1}
\makeatletter
\renewcommand{\theequation}{S\arabic{equation}}
\renewcommand{\thefigure}{S\arabic{figure}}

In this Supplementary Material we study the validity of the mean-field approximation, prove the compact relation between acceleration and cavity field as the result of conservation of momentum, detailed calculations in short-time dynamics as well as long-time dynamics in red and blue cavity-detuned regime, and with a three-level Dicke-like model. 

\section*{A --- Atom-Only Master Equation and validity of mean-field approximation}
We proof the validity of this approximation by deriving the atom-only master equation of this system. To start with, we divide the operator \(O=\sum_q c_{q+k}^\dag c_q\) into two parts, its expectation value and quantum fluctuation: \(O=\langle O \rangle +\delta O\). Then, the Lindblad master equation of the atom and cavity system is
\begin{equation}
\begin{aligned}
    \frac{d}{dt}\rho=&-i[-\Delta_c a^\dag a+\eta_p (\langle O\rangle a^\dag+\langle O\rangle^* a), \rho ]+\kappa(2a\rho a^\dag -a^\dag a\rho-\rho a^\dag a)\\
    & -i[\sum_q \frac{\hbar \omega_q}{2m} c_q^\dag c_q + \eta_p(\delta O a^\dag +\delta O^\dag a), \rho].
\end{aligned}
\end{equation}
The first line describe the dissipation-driven dynamics of the cavity field. In the time scale of cavity field dynamics, the change in \(\langle O\rangle\) is neglectable. Thus, \(\langle O \rangle\) acts as a quasi-static classical driving source on the cavity field. The second line describes the dynamics of atom and the interactions between cavity and fluctuations in atom operator \(\delta O\).

Since the atoms are coupled to a cavity with strong dissipation, the cavity field can be adiabatically eliminated and we get the atom-only master equation \cite{azouit2016adiabatic},
\begin{equation}
\begin{aligned}
    \frac{d}{dt}\rho_A=&-i[\sum_q \frac{\hbar \omega_q}{2m} c_q^\dag c_q+\eta_p(\alpha_0 O^\dag + \alpha_0^* O),\rho_A]+\frac{\eta_p^2}{\kappa}(2\delta O \rho_A \delta O^\dag - \delta O^\dag \delta O \rho_A-\rho_A\delta O^\dag \delta O),
\end{aligned}
\label{atom_lme}
\end{equation}
together with the relation between steady cavity field and  atomic order, \(
    \alpha_0=\langle a \rangle =\frac{\eta_p}{\Delta_c+i\kappa}\langle O\rangle\).
The first term in the atom-only master equation describes the coherent evolution of atoms driven by a self-generated classical light field. The second term describes the dissipative dynamics: the quantum fluctuation \(\delta O\) decays at rate \(\eta_p^2/\kappa\) and thus the coherent, mean field dynamics dominates.

\section*{B --- Acceleration and Conservation of momentum}
In this part, we connect the acceleration of the center-of-mass of the atoms with the cavity photon occupation. The total momentum of the atomic system is \(P=\sum_q q c^\dag_q c_q\). To calculate its evolution, we start with the equation of motion of the atomic operators from the main text,
\begin{equation}
i\partial_t c_q=\frac{\hbar q^2}{2m}c_q+\eta_p( a^\dag c_{q-k}+ac_{q+k}).
\label{AtomEvolution}
\end{equation}
Then, the time evolution of operator \(P\) is governed by
\begin{equation}
\begin{aligned}
i\partial_t P&=i\sum_q q(\partial_t c^\dagger_q \cdot c_q + c^\dagger_q\cdot\partial_t c_q)=\sum_q q\{[-\frac{\hbar q^2}{2m}c^\dagger_q-\eta_p(a c^\dagger_{q-k}+a^\dagger c_{q+k}^\dag)]c_q+c^\dagger_q[\frac{\hbar q^2}{2m}c_q+\eta_p( a^\dag c_{q-k}+ac_{q+k})]\}\\
&=\eta_p\sum_q q[a^\dag (c^\dag_qc_{q-k}-c^\dag_{q+k}c_{q})+a (c^\dag_qc_{q+k}-c^\dag_{q-k}c_{q})]
=\eta_pk(a^\dag O-aO^\dag),
\end{aligned}
\label{kEOMc1d}
\end{equation}
where \(O=\sum_q c^\dag_{q+k}c_q\). If quantum fluctuations are neglected, the expectation for the  acceleration is \begin{equation}
    \partial_t \langle P\rangle \approx -i\eta_p k (\langle a^\dag \rangle\langle O \rangle+\langle a \rangle\langle O^\dag \rangle).
\end{equation}
Finally, we adiabatically eliminated cavity field, \(\langle a \rangle\approx\eta_p\langle O \rangle/(\Delta_c+i\kappa)\), and get the compact relation
between acceleration and cavity field,
\begin{equation}
    \partial_t \langle P \rangle \approx \frac{(2\kappa)\eta_p^2 k}{\Delta_c^2+\kappa^2}|\langle O \rangle|^2 \approx (2\kappa) k |\langle a \rangle|^2.
    \label{MomentumConservation}
\end{equation}
In this equation, \(|\langle a \rangle|^2\) is the number of photon in the cavity, while \(2\kappa|\langle a \rangle|^2\) is the leaking rate of photons from the cavity. Since the relaxation of the cavity field is much faster then the atom dynamics, the number of photons leaking from the cavity is same as the number of photons scattered into the cavity in the time scale of atom dynamics. So, \((2\kappa) k |\langle a \rangle|^2\) is the momentum of the photons scattered by the atoms per unit time, which is equal to the momentum gained by the atoms per unit time according to the conservation of momentum.

\section*{C --- Short-time dynamics of atoms prepared in a pure momentum state}
\begin{figure*}
\includegraphics[scale=0.53]{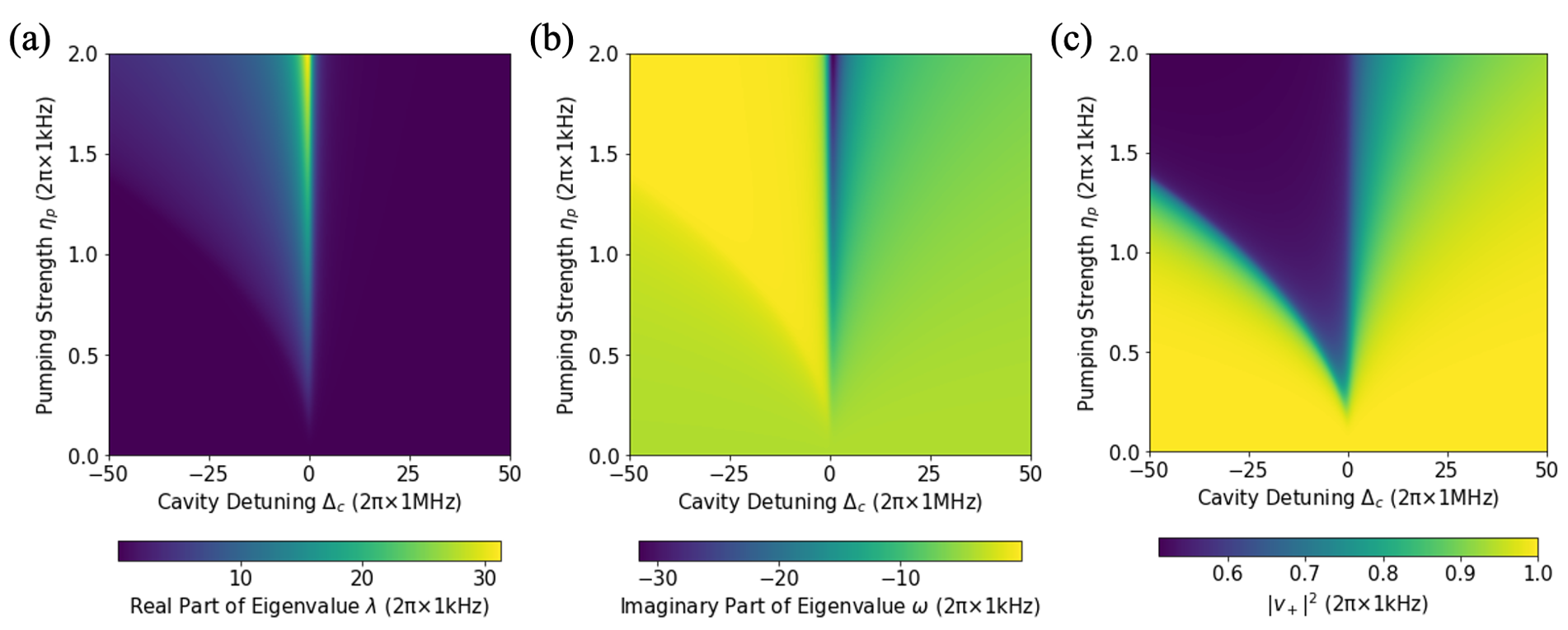}
\caption{\label{fig:S1} The eigenvalues and eigenvectors of the matrix \(M\) correspond to the dynamics. (a) The real part of the eigenvalue which is the exponential growth rate \(\lambda >0\) of the eigenmode. (b) The imaginary part of the eigenvalue which is the oscillation frequency \(\omega\) of the corresponding eigenmode. (c) The weight of \(q+k\) momentum component in the eigenvector \(|v_+|^2\).}
\end{figure*}
To understand how the transport start, as well as the difference between red and blue detuning regime, we discuss the short-time dynamics of atoms that are prepared in a pure momentum state:
\begin{equation}
    | \psi (t=0)\rangle=|\sqrt{N}_q\rangle=e^{\sqrt{N}(c_q^\dag-c_q)}|0\rangle.
\end{equation}
During a very short time (\(t\ll (\frac{N\eta_p^2}{|\Delta_c-i\kappa|})^{-1}\)) after the coupling is turned on, the atoms are only pumped into \(q-k\) and \(q+k\) state, since only these two momentum states are directly coupled to \(q\) state. Moreover, we neglect the depletion in \(q\) state and replace operator \(c_q\) with its expectation:
\begin{equation}
    c_q\to \langle c_q\rangle \approx \sqrt{N},\quad c_q^\dag\to \langle c_q^\dag\rangle \approx \sqrt{N}.
\end{equation}
With these two approximations together as well as adiabatic elimination of cavity mode, we obtain equation of motion with only two modes, \(q-k\) and \(q+k\):
\begin{equation}
    \partial_t 
    \left(\begin{array}{c}
        \langle c_{q+k} \rangle  \\
        \langle c_{q-k}^\dag \rangle 
    \end{array}\right)
    =M\left(\begin{array}{c}
        \langle c_{q+k} \rangle   \\
        \langle c_{q-k}^\dag \rangle  
    \end{array}\right)
    ,\quad M=-i
    \left(\begin{array}{cc}
        (\omega_{q+k}-\omega_{q})+\frac{N\eta_p^2}{\Delta_c-i\kappa} & \frac{N\eta_p^2}{\Delta_c-i\kappa}   \\
        -\frac{N\eta_p^2}{\Delta_c-i\kappa} & -(\omega_{q-k}-\omega_{q})- \frac{N\eta_p^2}{\Delta_c-\ii\kappa}  
    \end{array}\right).
\label{2modemodel}
\end{equation}
If the atoms are initially trapped in a static potential (\(q=0\)), \(\omega_{q+k}-\omega_{q}=\omega_{q-k}-\omega_{q}=\omega_r\).

The short-time dynamics of atoms in \(q-k\) and \(q+k\) modes is governed by the eigenvalues and eigenvectors of matrix \(M\) in Eqn. \eqref{2modemodel} should be calculated: 
\begin{equation}
    M 
    \left(\begin{array}{c}
        v_+  \\
        v_- 
    \end{array}\right)
    =(\lambda+i\omega)\left(\begin{array}{c}
        v_+  \\
        v_-  
    \end{array}\right).
\end{equation}
The eigenvector \([v_+,v_-]^T\) describes how the \(q+k\) and \(q-k\) states hybridize into the short-time eigenmode. The imaginary part \(\omega\) is the angular frequency of the eigenmode oscillation, while the real part \(\lambda\) is the rate exponential growth (\(\lambda>0\)) or decay (\(\lambda<0\)) rate of the eigenmode. Since we are not interested in the eigenmode that decays,  we focus on the eigenmode with positive real part eigenvalue (\(\lambda>0\)). In this eigenvector, \(q+k\) mode always dominates: \(|v_+|>|v_-|\). In the experiment, the atoms are often loaded in a static trap. So, we focus on the initial state \(q=0\) now. The real and imaginary part of the eigenvalue, and the weight of \(k\) momentum mode in the eigenvector \(|v_+|^2\) are plotted in the \((\Delta_c, \eta_p)\) parameter plane  (see Fig. \ref{fig:S1}). 

\begin{figure*}
\includegraphics[scale=0.53]{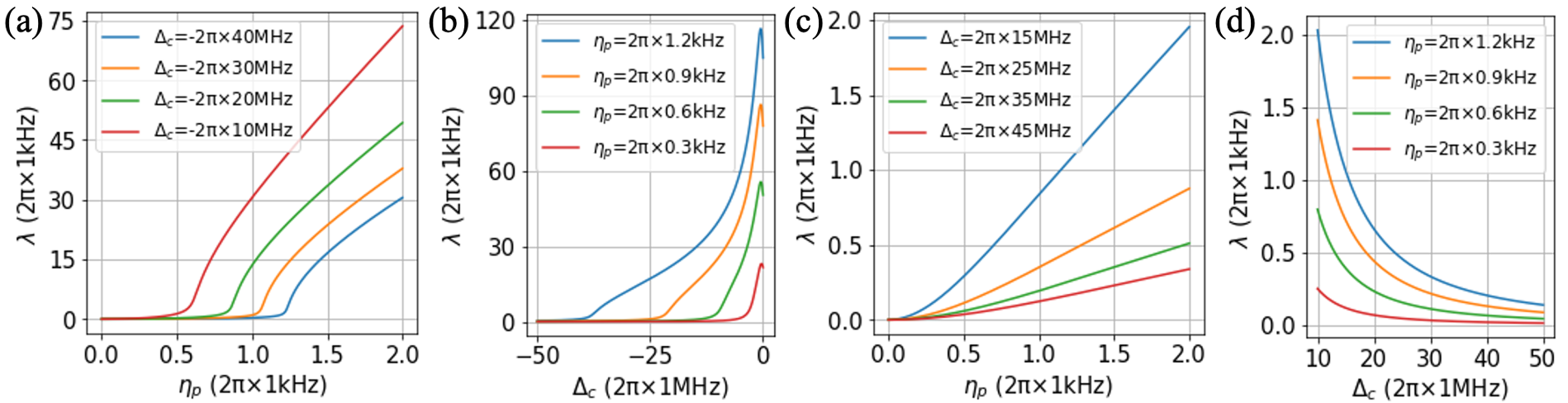}
\caption{\label{fig:S2} The exponential growth rate \(\lambda >0\) (a) as a function of pumping strength \(\eta_p\) at several fixed cavity detuning \(\Delta_c\) and (b) as a function of cavity detuning \(\Delta_c\) at several pumping strength \(\eta_p\) in the red cavity detuned regime; the exponential growth rate \(\lambda >0\) (c) as a function of pumping strength \(\eta_p\) at several fixed cavity detuning \(\Delta_c\) and (d) as a function of cavity detuning \(\Delta_c\) at several pumping strength \(\eta_p\) in the blue cavity detuned regime.}
\end{figure*}

In both red and blue cavity-detuned regime, there is no normal-to-superradiance phase transition. It seems that in the red regime, there is a critical boundary: the regime the system is in superradiance phase when \(\eta_p>\eta_p^{critical}(\Delta_c)\) and in normal phase when \(\eta_p<\eta_p^{critical}(\Delta_c)\). However, it is not a phase transition, since at any finite pumping strength \(\eta_p\), the exponential growth rate \(\lambda\) is positive (see Fig.\ref{fig:S2}). So, even at an infinitesimal pumping strength, the atoms will be pumped into the momentum mode \(k\) and \(-k\) and then spatial modulation in atom density and finite cavity field occurs. 

In the blue detuned regime, the eigenmode that atoms are pumped into is almost the \(k\) mode when the pumping is not very strong, since the weight of \(k\) mode in the eigenmode \(|v_+|^2\) is close to 1. So, the transition from \(q\) to \(q+k\) is enhanced and to \(q-k\) is suppressed in the blue detuned regime.

\section*{D --- Bloch wavefunction ansatz in the red-detuned regime}
We study the system in the regime $\Delta_c<0$ using as ansatz a Bloch wave function modulated by an envelope function (such as a Gaussian wave packet): 
\begin{equation}
    \phi(x)=f(x-x_0)e^{iqx}u_q(x-x_0),
    \label{bloch}
\end{equation}
Here, \(x_0\) is the center of the wave packet, \(q\) is the quasi-momentum, and \(u_q(x)\) is an even periodic function that satisfies \(u_q(x)=u_q(x+2\pi/k)\). The slow-varying envelope function \(f(x)\) satisfies \(|(\partial_x f(x))/f(x)|\ll k\), and we assume that the maximum of \(u_q(x)\) is located at \(x=0\). We adiabatically eliminate the fast evolving cavity field and find as steady state cavity field amplitude \(\alpha_0\):
\begin{equation}
    \begin{aligned}
    \alpha_0&=\frac{N\eta_p}{\Delta_c+i\kappa}\int dx \phi^*(x) e^{ikx}\phi(x)=\frac{N\eta_p}{\Delta_c+i\kappa}\ e^{ikx_0}\int |f(x)|^2 e^{ikx}|u_q(x)|^2dx\\
    &\approx \frac{N\eta_p}{\Delta_c+i\kappa}\ e^{ikx_0} \sum_{s=-\infty}^\infty |f(\frac{2\pi s}{k})|^2 \int_{\frac{\pi}{k}(2s-1)}^{\frac{\pi}{k}(2s+1)}e^{ikx}|u_q(x)|^2dx\approx \frac{N\eta_p}{\Delta_c+i\kappa}\ e^{ikx_0}\cdot \frac{k}{2\pi} \int_{-\frac{\pi}{k}}^{\frac{\pi}{k}}\cos(kx)|u_q(x)|^2dx.
    \end{aligned}
\end{equation}
Since integral in the last line is real, only the \((\Delta_c+i\kappa)^{-1}\) term contributes to the phase of \(\alpha_0\). Then, the optical potential of \(\alpha_0\) is
\begin{equation}
        V=\eta_p N|\alpha_0|\cos[kx-\arg(\alpha_0)]=\eta_p N|\alpha_0|\cos[k(x-x_0)-\arg(\frac{1}{\Delta_c+i\kappa})].
\end{equation}

In the red cavity detuned regime (\(\Delta_c<0\)), if the dissipation \(\kappa\) is nonzero, there is a phase shift \(\delta=\tan^{-1}(\kappa/|\Delta_c|)\) between atom wave function and the potential it generates. This deviation of atom wavefunction from the minima of potential causes an effective force on the atoms and lead to transport in \(+x\) direction. The expectation of the force on atoms to the first order approximation by neglecting the shape of \(u(x)\) is
\begin{equation}
        \langle \frac{dP}{dt}\rangle =\langle -\frac{\partial V}{\partial x}\rangle \approx -\frac{\partial V}{\partial x}\mid_{x=x_0}
        =-\eta_p Nk|\alpha_0|\sin(0-\delta)
        = \eta_p N k|\alpha_0|\frac{\kappa}{\sqrt{|\Delta_c|^2+\kappa^2}}.
\end{equation}
Together with Eqn.\eqref{MomentumConservation}, we solve the steady cavity field and effective force on atom (also the acceleration), as functions of experimental tunable parameters \(\eta_p\) and \(\Delta_c\), in the small phase shift (\(\Delta_c\ll-\kappa<0\)) regime:
\begin{equation}
    |\alpha_0|\approx \frac{N\eta_p}{2\sqrt{|\Delta_c|^2+\kappa^2}},\quad
    \langle \frac{dP}{dt} \rangle \approx\frac{\kappa k}{2}\frac{(N\eta_p)^2}{|\Delta_c|^2+\kappa^2}.
\end{equation}

\section*{E --- Self-Induced Rabi Oscillation in the blue-detuned regime}
\begin{figure*}
\includegraphics[scale=0.53]{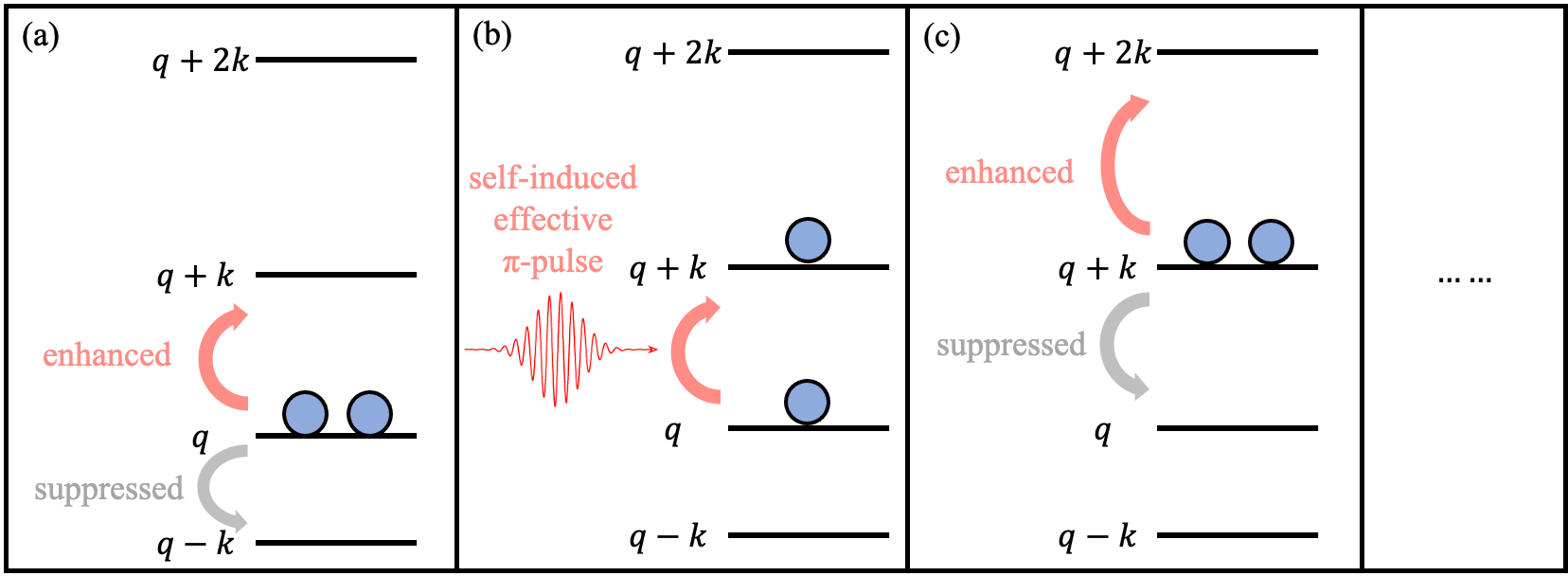}
\caption{Step-like acceleration of atoms in the blue-detuned regime. (a) Atoms are initialized in momentum mode $k$. Transitions to mode $k+q$ is enhanced, while transition to mode $k-q$ it is suppressed. (b) The transition from mode \(q\) to mode $k+q$ is driven by the cavity field generated by the spatially modulation of atoms as a result of the interference of these to momentum modes. (c) The process repeats from state $k+q$ to higher momentum states.  \label{fig:S3} }
\end{figure*}

The transport phenomena of atoms in the blue cavity detuned and weak pumping regime can be understood as self-induced Rabi oscillation between two neighbour momentum states \(q\) and \(q+k\). Support the atoms are initialized in momentum mode \(q\). In the blue regime, the transition from \(q\) state to \(q-k\) state is suppressed and transition to \(q+k\) state is enhanced by the dissipation. So, only \(q\) and \(q+k\) are occupied and involved in the dynamics. The equations of motion of these two states are
\begin{equation}
    i\partial_t 
    \left(\begin{array}{c}
        \langle c_{q} \rangle  \\
        \langle c_{q+k} \rangle 
    \end{array}\right)
    =
    \left(\begin{array}{cc}
        \omega_{q} & \eta_p \alpha   \\
        \eta_p \alpha^* & \omega_{q+k}   
    \end{array}\right)
    \left(\begin{array}{c}
        \langle c_{q} \rangle   \\
        \langle c_{q+k} \rangle  
    \end{array}\right).
\end{equation}
The field amplitude is \(\alpha=\frac{N\eta_q}{\Delta_c+i\kappa}\langle c_{q+k}^\dag\rangle\langle c_q\rangle\) under mean field approximation. If the coupling \(\eta_p \alpha\) is weak, the amplitude \(\langle c_q\rangle\) and \(\langle c_{q+k}\rangle\) can be separated into a slow-varying term and a free oscillation term:
    \(\langle c_{q} \rangle =\Tilde{c}_q(t) e^{-i\omega_q t}\), \(\langle c_{q+k} \rangle =\Tilde{c}_{q+k}(t)e^{-i\omega_{q+k} t}\). Then, the cavity field is
\begin{equation}
    \alpha (t)=\frac{N\eta_q}{\Delta_c+i\kappa}\Tilde{c}_{q+k}^*(t)\Tilde{c}_q(t)e^{i(\omega_{q+k}-\omega_q)t},
\end{equation}
which is a classical light which is on-resonance to the transition frequency between \(q\) and \(q+k\) level, and modulated by a pulse with shape \(\Tilde{c}_{q+k}^*(t)\Tilde{c}_q(t)\). It drives a "Rabi Oscillation" between \(q\) and \(q+k\) mode. In the rotating frame, the equation of motion is
\begin{equation}
    i\partial_t 
    \left(\begin{array}{c}
        \Tilde{c}_{q}   \\
        \Tilde{c}_{q+k}
    \end{array}\right)
    =
    \left(\begin{array}{cc}
        0 & \frac{N\eta_q^2}{\Delta_c+i\kappa}\Tilde{c}_{q+k}^*\Tilde{c}_q   \\
        \frac{N\eta_q^2}{\Delta_c-i\kappa}\Tilde{c}_{q}^*\Tilde{c}_{q+k}  & 0   
    \end{array}\right)
    \left(\begin{array}{c}
        \Tilde{c}_{q}   \\
        \Tilde{c}_{q+k}  
    \end{array}\right).
\end{equation}
Once a "\(\pi\)-pulse" is applied to the \(q\) and \(q+k\) two level system, all atoms are coherently driven into \(q+k\) mode from \(q\) mode. The occupation of \(q\) mode drops to zero and the driving between these two level \(\Tilde{c}_{q+k}^*(t)\Tilde{c}_q(t)\) is automatically turned off. Then, the \(q+k\) state is occupied by all atoms and a new cycle of transition from \(q+k\) to \(q+2k\) mode starts. Fig. \ref{fig:S3} is a illustration of this process that atoms climb the momentum ladder and accelerate in a step-like fashion.

During the "Rabi oscillation" between \(q\) and \(q+k\) mode, both modes are occupied. Thus, transition from \(q+k\) to \(q+2k\) state may simultaneously happen. The detuning between the frequency of driving generated by atoms in \(q\) and \(q+k\) mode, and the level seperation between \(q+2k\) and \(q+k\) mode is
\begin{equation}
    \delta_q=|\omega_{drive}-(\omega_{q+2k}-\omega_{q+k})|=|[(q/k+1)^2-(q/k)^2]-[(q/k+2)^2-(q/k+1)^2]|\omega_r=2\omega_r.
\end{equation}
If the driving is much smaller than the level detuning \(\delta_q\):
\begin{equation}
    \eta_p\alpha<\frac{N\eta_q^2}{4|\Delta_c+i\kappa|}\ll \delta_q=2\omega_r,
\end{equation}
then the leakage is neglectable. If the pumping \(\eta_p\) is strong enough or the cavity deturning \(\Delta_c\) is too small, the transitions to the momentum states higher than \(q+k\) are not neglectable. Once many momentum modes are sufficiently occupied, the components in cavity field that oscillating at the frequency difference between any two modes occur. These different frequency components cause dephasing in cavity field and the picture of self-induced Rabi oscillation of two levels breaks down. 

\section*{F --- The Three-level Dicke-like model}

We begin by taking the Hamiltonian from the main text,
\begin{equation}
    H=-\hbar \Delta_c a^\dag a+\sum_q \hbar \omega_q c^\dagger_q c_q+\eta_p \left(a^\dag \left[\sum_q c^\dagger_{q+k}c_q\right]+h.c.\right).
\end{equation}
We reduce to the three mode approximation, i.e. we keep only the $c_1:=c_q$ and $c_2:=c_{q+k}$, $c_3:=c_{q-k}$ modes in the Hamiltonian. We may then perform a $SU(3)$ Schwinger boson mapping \cite{3Dicke1}, $S^+_{12}=c^\dagger_{2}c_1$, $S^+_{13}=c^\dagger_{3}c_1$, $S^-_{\mu}=(S^+_{\mu})^\dagger$, $S^z_{12}=c^\dagger_{2}c_{2}-c^\dagger_{3}c_{3}-c^\dagger_{1}c_{1}$, $S^z_{13}=c^\dagger_{3}c_{3}-c^\dagger_{2}c_{2}-c^\dagger_{1}c_{1}$, which obey the $SU(3)$ algebra with representation $N$ \cite{lacroix2011introduction,3Dicke1} enforcing the total atom number $N=c^\dagger_{3}c_{3}+c^\dagger_{2}c_{2}+c^\dagger_{1}c_{1}$. This allows us to rewrite the model, up to an irrelevant constant, in terms of a three-level Dicke-like one,
\begin{equation}
H=\omega_0 (S^z_{12}+S^z_{13})+\Delta_c a^\dag a+\eta_p \left(a^\dagger (S^+_{12}+S^-_{13})+h.c.\right), \label{TC}
\end{equation}
where $\omega_0=\frac{\omega_{q+k}-\omega_q}{2}$. We now switch to the large atom number limit $N \to \infty$ and perform a generalized Holstein-Primakoff transformation \cite{3Dicke1,wagner1975nonlinear} to obtain,
\begin{equation}
H=\omega_0 (N_{12}+N_{13})+\Delta_c a^\dag a+\eta_p \left(a^\dagger (a^\dagger_{12}+a_{13})+h.c.\right), \label{bosonH}
\end{equation}
where we defined $a_x$ to be bosonic operators and $N_x=a^\dagger_x a_x$. The full model including the Lindblad jump operator is now quadratic and may exactly solved \cite{Prosen_2008,prosen2010quantization}. It is simplest to write the equations of motion for one-point functions $\vec{v}=\{\ave{a},\ave{a}^*,\ave{a_{12}},\ave{a_{12}}^*,\ave{a_{13}},\ave{a_{13}}^*\}$. These equations, of course, close into a set of six equations. Using Eqn. \eqref{MasterEqn} we may find,
\begin{equation}
\frac{d \vec{v}}{dt}= {\cal M} \vec{v},
\end{equation}
where,
\begin{equation}
{\cal M}=\left(
\begin{array}{cccccc}
 \ii \Delta _c-\kappa  & 0 & 0 & \ii \eta _p & -\ii \eta _p & 0 \\
 0 & -\kappa -\ii \Delta _c & -\ii \eta _p & 0 & 0 & i \eta _p \\
 0 & \ii \eta _p & -\ii \omega _0 & 0 & 0 & 0 \\
 -\ii \eta _p & 0 & 0 & \ii \omega _0 & 0 & 0 \\
 -\ii \eta _p & 0 & 0 & 0 & -\ii \omega _0 & 0 \\
 0 & \ii \eta _p & 0 & 0 & 0 & \ii \omega _0 \\
\end{array}
\right)
\end{equation}
The eigenvalue equation $\det({\cal M}-\Lambda)=0$ for this matrix is,
\begin{equation}
\Delta _c^2 \left(\Lambda ^2+\omega _0^2\right)^2+4 \omega _0 \Delta _c \left(\Lambda ^2+\omega _0^2\right) \eta _p^2+(\kappa +\Lambda )^2 \left(\Lambda ^2+\omega _0^2\right)^2+4 \omega _0^2 \eta
   _p^4=0
\end{equation}
Assuming $\kappa \to \infty$ as the experimentally relevant bad cavity limits we may expand this equation in $1/\kappa$. We always have two eigenvalues with positive real part (depending on the sign of $\Delta_c$) $\Lambda_{{\rm div}}=\{\frac{1}{6} \left(\sqrt{3}+\ii\right)\Delta_c,\frac{1}{6} \left(-\sqrt{3}+\ii\right)\Delta_c\}$ implying that the model is always unstable. This is consistent with the numerics which implies that we cannot truncate to a finite number of modes in the long-time limit. However, we also see that at the point $\Delta_c=0$ the two unstable solutions interchange signaling the phase transition between the red and blue detuned regime from the numerical study in the main text. This also suggests that the phase transition is a manifestation of quantum Zeno dynamics \cite{Zeno1,bezvershenko2020dicke}. Indeed, setting $\kappa$=0, by looking at linear stability of analysis of the trivial normal phase \cite{Dickereview}, we find a standard Dicke phase transition between a normal and superradiant phase, further confirming the dissipation-induced nature of our spatio-temporal lattice. Crucially, in our model, due to the mapping at momentum $q$ and $q+k$, the oscillations are accompanied by formation of a spatial lattice, as well as the temporal one (persistent oscillations). 

\end{document}